\documentclass[superscriptaddress,secnumarabic,
amssymb,amsmath,nobibnotes,aps,prd,showkeys,showpacs,nofootinbib]{revtex4}%
\pdfoutput=1
\usepackage{graphicx}
\usepackage{epsf}
\usepackage{epsfig}
\usepackage{scrtime}
\usepackage[multiple]{footmisc}
\usepackage{amsfonts, graphics}
\usepackage{bm}
\usepackage[utf8x]{inputenc}
\usepackage{amsmath}
\usepackage{amsfonts}
\usepackage{amssymb}
\usepackage{multirow}
\usepackage{epstopdf}
\usepackage{natbib}%
\providecommand{\U}[1]{\protect\rule{.1in}{.1in}}
\newcommand{\be}{\begin{equation}}
\newcommand{\ee}{\end{equation}}

\newcommand{\mincir}{\raise
-3.truept\hbox{\rlap{\hbox{$\sim$}}\raise4.truept\hbox{$<$}\ }}
\newcommand{\magcir}{\raise
-3.truept\hbox{\rlap{\hbox{$\sim$}}\raise4.truept\hbox{$>$}\ }}

\begin{document}
\title{ $f(R)$ gravity solutions for Evolving Wormholes }
\author{Subhra Bhattacharya}
\email{subhra.maths@presiuniv.ac.in}
\affiliation{Department of Mathematics, Presidency University, Kolkata-700073, India}
\author{Subenoy Chakraborty}
\email{schakraborty@math.jdvu.ac.in}
\affiliation{Department of Mathematics, Jadavpur University, Kolkata-700032, India}

\begin{abstract}

Scalar-tensor $f(R)$ theory of gravity is considered in the framework of a simple inhomogeneous space-time model. In this we use the reconstruction technique to look for possible evolving wormhole solutions within viable $f(R)$ gravity formalism. These $f(R)$ models are then constrained so that they are consistent with existing experimental data. Energy conditions related to the matter threading the wormhole are analysed graphically and are in general found to obey the null energy conditions (NEC) in regions around the throat, while in the limit $f(R)=R,$ NEC can be violated at large in regions around the throat.
\end{abstract}
\maketitle

Wormhole is a hypothetical object which acts like a bridge or tunnel to connect asymptotic regions of a single universe or two distinct universes. This imaginary, intuitive concept is one of the most popular and intensively studied area of research in general relativity. The nature of wormholes studied so far can be classified as static wormholes and dynamic wormholes. Although there are wormhole solutions \cite{ellis1,ellis2,bronnikov,kodama,clement} in the literature since 1973, but pioneering work related to static wormholes is due to {\it Morris and Thorne} \cite{m+t}. Such a wormhole is characterized by the metric: 
\begin{equation}ds^{2}=-e^{2\Phi(r)}dt^{2}+\frac{dr^{2}}{1-b(r)/r}+r^{2}(d\theta^{2}+\sin^{2}\theta d\phi^{2})\label{MTM}
\end{equation} with $c=G=1.$ Here $\Phi(r)$ is called the red-shift function. This red shift function help us identify the event horizon, since $g_{00}= -e^{2\Phi}\rightarrow 0$ on the event horizon. The function $b(r)$ is called the shape function and indicative of the spatial shape of the wormhole. Normally, it is seen that wormhole geometries are not dictated by Einstein field equations, rather the red shift and shape function (i.e. space-time metric) are chosen by hand and then one uses the Einstein field equations to evaluate the corresponding matter part. The matter part so obtained satisfies the local conservation equation due to Bianchi identities and violates the NEC \cite{visserb, visser1, visser2,ida, fewster, zavalsk,visser5}. However for asymptotically flat space time the violation of NEC is due to topological censorship \cite{frdmn,galloway}. These wormholes are considered traversable as they can facilitate two way passage through them. For a wormhole to be traversable, $e^{2\Phi(r)}$ should be finite everywhere (no horizon condition) and at the throat $r=r_{0}$ the condition $b(r_{0})=r_{0}$ holds. Since superluminal travel is possible through such a wormhole, they can be conceived of as hypothetical time machines \cite{visserb,visser3,lobo1,m+t+y,thorne,visser4}.

Generalizing the original Morris-Thorne metric (\ref{MTM}) by inducting the scale factor $a(t)$ gives us the evolving relativistic wormholes. Evolving relativistic wormholes or rather dynamical wormholes are not very popular in the literature, and hence there is only limited understanding on the topic.  Various different work on dynamical wormholes speculated their existence with matter satisfying WEC and DEC \cite{kar1,kar2,harada1,harada2}. While, there exists other works \cite{visser6,hay1} speculating that matter threading the wormhole would violate NEC. Recently dynamical wormholes have been treated with two fluid system in \cite{catl2, catl3}, important from the perspective of accelerated expansion of the universe \cite{catl4,bani,cai,pan1} and also in \cite{pan2} with dissipation due to particle creation mechanism in non-equilibrium thermodynamics.  

In this work we extend the analysis of dynamic wormholes in scalar-tensor $f(R)$ modified theories of gravity. Traversable wormholes have already been treated in modified gravity (for example in Einstein-Gauss-Bonnet gravity, in higher dimensional Lovelock theories \cite{dotti1, dotti2, dotti3, matulich, m+n}, in $f(R)$ gravity theories \cite{harko, furey} and in brane scenario like the Randall-Sundram brane \cite{la}) without any exotic matter. However such work is limited for their dynamic counterparts. Scalar-tensor $f(R)$ gravity theory was developed by adopting a modification in the Einstein-Hilbert gravitational Lagrangean density involving an arbitrary function of the Ricci scalar $R$ \cite{frgravit} and essentially, it is an higher order $(4^{th})$ gravity theory. In this context it can be stated that, since Einstein's formulation of General Relativity, several extensions of it have been introduced but with quite disparate reasons. {\it Weil} \cite{weil} initiated such attempt with the motivation of unifying electromagnetic and gravitational interaction geometrically, also there were modifications, namely to the gravitational action in analogy to quantum field theory in curved space-time, to have fundamental unification schemes and to obtain complete geometric explanation of the dark sector phenomenology \cite{dono}. $f(R)$ gravity theory was probably one of the simplest such extension, and can act as natural model for inflation \cite{starobin}. Also it has been shown to act as a geometrical dark energy model \cite{drkeng}. Scalar-tensor $f(R)$ gravity theory has also been used to address the problem of dark matter and can be an alternative contributor of non-baryonic dark matter \cite{capo, cemb}. More recently standard $f(R)$ gravity theory has been used to explain the cosmic evolution since Big-Bang starting from early inflation to dark energy \cite{capo1}. Hence $f(R)$ gravity is an alternative framework based on general relativity that does not need any new scalar field to explain the accelerated expansion of the universe. 
The action in $f(R)$ theory is given by:        
\begin{equation}
A=\frac{1}{2\kappa}\int\sqrt{-g}f(R)d^{4}x+\int d^{4}x\sqrt{-g}\mathcal{L}_{m}(g_{\mu \nu}\Psi)\label{action}
\end{equation}
where $\kappa=8\pi G,~R$ is the usual Ricci Scalar, $f(R)$ is an arbitrary function of $R$ and $\mathcal{L}_{m}$ is the matter Lagrangian density with the corresponding integral denoting matter action with metric $g_{\mu\nu}$ coupled with collective matter field $\Psi$. Now variations of the above action with respect to the metric coefficient gives the fourth order field equations
\begin{equation}
F(R)R_{\mu}^{\nu}-\frac{1}{2}f(R)\delta_{\mu}^{\nu}-\nabla_{\mu}\nabla^{\nu}F(R)+\delta_{\mu}^{\nu}\Box F(R)=T_{\mu}^{\nu}\label{fieldeqn}
\end{equation}
where $R_{\mu}^{\nu}$ is the Ricci tensor, $T_{\mu}^{\nu}$ is the stress energy tensor of matter part and  $F(R)=\frac{\partial f}{\partial R}.$ Here we are interested in constructing dynamical wormholes having viable $f(R)$ gravity solutions within the framework of standard cosmological assumptions. We use the classical reconstruction technique to obtain $f(R)$ solutions. Reconstruction technique has been used in the several context to get viable $f(R)$ gravity forms mimicking the expansion history of the universe \cite{dom, carl, dunb, noj, cog1}. In this work we follow a method similar to that used in \cite{lobo} and assume some specific forms of the shape function and scale factor so as to construct the $f(R)$ gravity solutions with prescribed matter content from the field equations. We consider the spherically symmetric inhomogeneous space-time metric given by
\begin{equation}
ds^{2}=-dt^{2}+a^{2}(t)\left[\frac{dr^{2}}{1-b(r)/r}+r^{2}(d\theta^{2}+\sin^{2}\theta d\phi^{2})\right].\label{metric}
\end{equation} It is evident from the metric that, we have considered the red shift function $\Phi(r)=constant,$ { \it i.e.} we are considering zero-tidal force wormhole solutions. For homogeneous FRW model it is now known that zero tidal force wormholes are supported by anisotropic fluid \cite{catl}. Although there is no such general result for the inhomogeneous model considered, still we have considered them because of their mathematical tractability and simplicity, which makes them altogether interesting to study important properties of wormhole geometry.

Now taking trace of the field equations (\ref{fieldeqn}), (after simple algebraic manipulations) we get
 \begin{equation}
 G_{\mu\nu}=\frac{1}{F}T_{\mu\nu}+\frac{1}{F}[\nabla_{\mu}\nabla_{\nu}F-g_{\mu\nu}N]\label{fieldeqn1}
 \end{equation}
 with \begin{equation}
 N(t,r)=\frac{1}{4}(RF+\Box F+T).\label{N}
\end{equation}  
and
\begin{equation}
 T=T_{\mu}^{\mu}.\label{T}
\end{equation} 
Here the matter threading the wormhole is anisotropic, $T=-\rho+p_{r}+2p_{t}$ and it satisfies the conservation equations $(T_{\mu;\nu}^{\nu}=0)$ given by
\begin{align}
 \frac{\partial \rho}{\partial t}+H(3\rho+p_{r}+2p_{t})=0\label{consrv1}\\
 \frac{\partial p_{r}}{r}-\frac{2}{r}(p_{t}-p_{r})=0\label{consrv2}
\end{align}
For the metric(\ref{metric}) we get the explicit form of the field equations (\ref{fieldeqn1}) as:
\begin{equation}
\left.\begin{aligned}
3H^{2}+\frac{b^{'}}{a^{2}r^{2}}=\frac{\rho}{F}+\frac{\rho_{g}(r,t)}{F}~~~~~~~~~~~~~~ \\
 -(2\dot{H}+3H^{2})-\frac{b}{a^{2}r^{3}}=\frac{p_{r}}{F}+\frac{p_{rg}(r,t)}{F} ~~\\
 -(2\dot{H}+3H^{2})+\frac{b-rb^{'}}{2a^{2}r^{3}}=\frac{p_{t}}{F}+\frac{p_{tg}(r,t)}{F}
 \end{aligned}
\right\}
\label{fieldeqn2}
\end{equation}
where an over dot indicates differentiation w.r.t. $t$ and prime indicates differentiation w.r.t. $r$, and $\rho=\rho(r,t),~p_{r}=p_{r}(r,t),~p_{t}=p_{t}(r,t)$ are respectively the energy density and thermodynamic radial and transverse pressure of the anisotropic fluid that should satisfy the conservation equations (\ref{consrv1}) and (\ref{consrv2}). Further $\rho_{g},~p_{rg}$ and $p_{tg}$ the energy density, radial and transverse pressure of a hypothetical curvature fluid satisfies the following:
\begin{equation}
\left.\begin{aligned}
\rho_{g}=N+\ddot{F}~~~~~~~~~~~~~~~~~~~~~~~~~~~~~~~~~~~~~  \\
p_{rg}=-N-H\dot{F}+\frac{r-b}{a^{2}r}F^{''}-\frac{b-rb^{'}}{2a^{2}r^{2}}F^{'}\\
p_{tg}=-N-H\dot{F}+\frac{r-b}{a^{2}r^{2}}F^{'}.~~~~~~~~~~~~~~~~~
\end{aligned}
\right\}
\label{rhovals}
\end{equation}
Note that the field equations (\ref{fieldeqn2}) can be interpreted as the Einstein field equations with non-interacting two-fluids. Now the scalar curvature for the space-time metric (\ref{metric}) has the expression \begin{equation}
R=6(\dot{H}+2H^{2})+\frac{2b^{'}}{a^{2}r^{2}}\label{R}
\end{equation}
{\it i. e.} $R=R(r,t).$ In order to find out expressions for $f(R)$ we shall make some specific choices of the shape function $b(r)$ as follows:

\vspace{1em}
{\it Case 1: $b=b_{0}\left(\frac{r}{r_{0}}\right)^{3}+d_{0}$ and $a=a_{0}\left(\frac{t}{t_{0}}\right)^{n}$:}
\vspace{1em}

For this choice of $b(r)$ and $a(t)$ we get
\begin{equation}
R=R(t)=\frac{6n}{t^{2}}(2n-1)+\frac{6b_{0}}{a^{2}r_{0}^{3}}\label{R1}
\end{equation}
 so that $F(R)$ is now a function of the $t$ co-ordinate alone, {\it i. e.} $F(R)=\psi(t).$ 
Now from the field equations (\ref{fieldeqn2}) and the conservation equations (\ref{consrv1}) and (\ref{consrv2}) we can get 
\begin{equation}
\left.\begin{aligned}
\rho=\frac{3n^{2}}{t^{2}}\psi(t)~~~~~~~~~~~~~~~~~~~~~~~~~~~~~~~~~~~~~~\\ 
p_{r}=\frac{n}{t^{2}}(2-3n)\psi(t)-\frac{d_{0}}{a^{2}r^{3}}\psi(t)-\frac{n}{t}\dot{\psi}(t)~\\ 
p_{t}=\frac{n}{t^{2}}(2-3n)\psi(t)+\frac{d_{0}}{2a^{2}r^{3}}\psi(t)-\frac{n}{t}\dot{\psi}(t)
\end{aligned}
\right\}
\label{p1}
\end{equation}

From equation (\ref{rhovals}) we get
\begin{equation}
\left.\begin{aligned}
\rho_{g}=\frac{3b_{0}}{a^{2}r_{0}^{3}}\psi(t) ~~~~~~~~~~~~~~~\\
p_{rg}=-\frac{b_{0}}{a^{2}r_{0}^{3}}\psi(t)+\frac{n}{t}\dot{\psi}(t) \\
p_{tg}=-\frac{b_{0}}{a^{2}r_{0}^{3}}\psi(t)+\frac{n}{t}\dot{\psi}(t)
\end{aligned}
\right\}
\label{pg1}
\end{equation}

Correspondingly the differential equation for $\psi$ takes the following form:
\begin{equation}
t^{2n}\ddot{\psi}-2nt^{2n-1}\dot{\psi}-\frac{2b_{0}t_{0}^{2n}}{a_{0}^{2}r_{0}^{3}}\psi=0.\label{DE1a}
\end{equation} 
This has a solution of the form:
\begin{equation}
\psi(t)=\left(\frac{2a_{0}^{2}r_{0}^{3}t_{0}^{2n}(1-n)^{2}}{b_{0}}\right)^{\frac{2n+1}{2(n-1)}}t^{\frac{2n+1}{2}}\Gamma\left(\frac{4n-1}{2n-2}\right)I_{\frac{2n+1}{2n-2}}\left[\frac{\sqrt{2b_{0}}t_{0}^{n}t^{(1-n)}}{a_{0}(1-n)r_{0}^{3/2}}\right],~n\neq 1.\label{solu1a}
\end{equation}$I_{\alpha}$ being the modified Bessel's function of the first kind. 

For two specific values of $n$ given by $n=1/2$ and $n=2/3$ we get the following solutions for $F(R)$ respectively
\begin{equation}
F(R)=\frac{2R_{0}}{3R_{1}^{1/2}}I_{2}\left(2\sqrt{\frac{R_{0}}{3R_{1}^{1/2}R}}\right)\label{Fr1}
\end{equation} 
and
\begin{equation}
F(R)=\left(1+\frac{12\mathcal{R}}{5}\right)\cosh\left(2\sqrt{\frac{\mathcal{R}}{3}}\right)-\sqrt{3\mathcal{R}}\left(\frac{8\mathcal{R}}{5}+3\right)\sinh\left(2\sqrt{\frac{\mathcal{R}}{3}}\right).\label{Fr2}
\end{equation} (It can be noted that although the above two solutions of $F(R)$ are chosen corresponding to the cosmological periods of radiation and matter dominated epoch of a homogeneous FRW universe, however one must keep in mind that it bears no physical significance from the perspective of the present inhomogeneous model). Here $R_{0}=\frac{6b_{0}}{r_{0}^{3}a_{0}^{2}},~R_{1}=\frac{1}{t_{0}^{2}}$ and $\mathcal{R}=\left\{\left[\frac{4R_{1}}{R}\right]\frac{[(12R_{1}^{2}R-R_{0})^{1/2}+(12R_{1}^{2}R)^{1/2}]^{4/3}+R_{0}^{2}}{[(12R_{1}^{2}R-R_{0})^{1/2}+(12R_{1}^{2}R)^{1/2}]^{2/3}}-1\right\}^{-1}.$

\vspace{1em}

For $n=1$ we have the simple power law form of solution given by
\begin{equation}
F(R)=\left(\frac{6R_{1}+R_{0}}{R_{1}R}\right)^{3/4}\left[C_{1}\left(\frac{6R_{1}+R_{0}}{R_{1}R}\right)^{\frac{m_{1}}{2}}+C_{2}\left(\frac{6R_{1}+R_{0}}{R_{1}R}\right)^{\frac{m_{2}}{2}}\right]\label{Fr3}
\end{equation} with $m_{1,2}=\pm\frac{3}{2}\sqrt{1+\frac{4R_{0}}{27R_{1}}}$

\vspace{1em}

In the above solutions we note that viable $f(R)$ solutions involving exponential of curvature $R$ have been explored in \cite{cog, lind} and have been successfully used to explain cosmological dynamics. Also solutions involving the negative powers of curvature may be obtained from some time-dependent compactification of string/M-theory \cite{nojiri}, further quantum fluctuations in nearly flat space-time may also include such terms \cite{vassi}. Also the negative power terms can serve as gravitational alternative to Dark Energy and produce cosmic acceleration \cite{sotirio,sotirio1}, while positive powers can induce early inflation. Moreover such modified gravity theories (of the type $R+R^{m}+\frac{1}{R^{n}}$ with $m,~n$ positive numbers) do not suffer from instabilities as compared to the ones given by $R+\frac{1}{R}$ which can suffer from instability. Also it is claimed that they might pass the solar system tests for scalar-tensor gravity, however more quantitative analysis is required to establish the relevance of such models in a real cosmological scenario \cite{nojiri1}. 
 
 \vspace{1em}
 
 {\it Special Case: $b_{0}=0$ i.e. we get $b(r)=d_{0}$}
 
 \vspace{1em}
 
 Substituting $b_{0}=0$ in equations (\ref{p1}) and (\ref{pg1}) we get the following equation:
 \begin{equation}
 \ddot{\psi(t)}-\frac{2n}{t}\dot{\psi(t)}=0,\label{DE1b}
 \end{equation}
 which has the solution:
 \begin{equation}
 F(R)=\frac{C_{1}}{1+2n}\left(\frac{R_{2}}{R}\right)^{\frac{2n+1}{2}}+C_{2},~n\neq-\frac{1}{2}\label{solu1b}
 \end{equation} where $R_{2}=6n(2n-1).$ As is evident in this case for $n= \frac{1}{2}$ we get $f(R)=R$ and then the system reduces to a case of Einstein's GR corresponding to the radiation epoch.
 
 For $n=-\frac{1}{2}$ we get the solution
 \begin{equation}
 F(R)=C_{1}\log\left(\sqrt{\frac{6}{R}}\right)+C_{2}\label{solu1c}
 \end{equation}
 
 Thus equations (\ref{solu1b}) and (\ref{solu1c}) represents the possible $F(R)$ gravity solutions corresponding to the evolving wormhole metric having constant throat radius $d_{0}.$ 
 
\vspace{1em}
 
 We shall consider another special case with $b(r)=constant$ for which we can get $f(R)$ gravity solution for some special values of $\rho,~p_{r}$ and $p_{t}.$
 
 \vspace{1em}
{\it Case 2: $b(r)=b_{0}$ and $a=a_{0}\left(\frac{t}{t_{0}}\right)^{n}$:}
\vspace{1em}

From the field equations we get:
\begin{equation}
\left.\begin{aligned}
\rho&=0\\
p_{r}&=-\frac{b_{0}}{a^{2}r^{3}}\psi(t)\\
p_{t}&=\frac{b_{0}}{2a^{2}r^{3}}\psi(t)
\end{aligned}
\right\}
\label{rhovals3}
\end{equation} 
which further gives the solution
\begin{equation}
\left.\begin{aligned}
\rho_{g}&=\frac{3n^{2}}{t^{2}}\psi(t)\\
p_{rg}&=-\frac{n(3n-2)}{t^{2}}\psi(t)\\
p_{tg}&=-\frac{n(3n-2)}{t^{2}}\psi(t)
\end{aligned}
\right\}
\label{rhovals4}
\end{equation} 
and $T=T_{\mu}^{\mu}=0.$ We note that conservation equations (\ref{consrv1}) and (\ref{consrv2}) are identically satisfied. Now using (\ref{rhovals4}) in (\ref{rhovals}) we get the following equation:

\begin{equation}
t^{2}\ddot{\psi(t)}-nt\dot{\psi(t)}-2n\psi(t)=0.\label{DE2}
\end{equation}
Hence for any $n$ we have the solution
 \begin{equation}
F(R)=\left(\frac{R_{2}}{R}\right)^{\frac{n+1}{4}}\left[C_{1}\left(\frac{R_{2}}{R}\right)^{-\frac{n_{1}}{4}}+C_{2}\left(\frac{R_{2}}{R}\right)^{-\frac{n_{2}}{4}}\right],~n>\frac{1}{2}\label{F21}
\end{equation}
with $n_{1,2}=\mp\sqrt{n^2+10n+1}$
and $R_{2}$ defined as before. Therefore we get $F(R)$ gravity solution for the above evolving wormhole with constant throat radius $b_{0},$ although matter here may not be considered real.

We note that such solutions for $f(R)$ has been previously obtained in \cite{sotirio, sotirio1, nojiri1}. As has been already stated such $f(R)$ gravity models containing both positive and negative powers of the scalar curvature can in fact describe both early time inflation and late time cosmic acceleration \cite{sotirio1, nojiri1}. Although here one must keep in mind that such an account is purely qualitative in nature and lacks any rigorous quantitative verification. 

Interestingly, this case has no physical significance because matter here is exotic (a consequence of the choice of solution), although the case is important mathematically considering viable $f(R)$ gravity solutions can exist in an otherwise unrealistic situation. 

\vspace{1em}
{\it Admissibility of the obtained $F(R)$ solutions:}
\vspace{1em}

 We shall now examine whether the present $f(R)$ models are viable or not i.e. whether the above $f(R)$ models are consistent with existing experimental and observational data.

It is commonly known that in metric formulation $f(R)$ gravity theory is equivalent to scalar-tensor gravity \cite{whitt,maeda,chiba,magnano} with coupling parameter $\omega=0$ \cite{chiba} (as in Brans-Dicke Theory). As solar system tests rule out small values of $\omega,$ so $f(R)$ theory is not a realistic one. However, it is known \cite{khoury,hu} that in strongly massive region, the non-minimally coupled scalar degree of freedom can be largely suppressed due to acquisition of excess mass and hence one cannot make such simple equivalence between the two gravity theories. Thus one should impose a set of constraints on $f(R)$ model to pass all of the tests. The following are the conditions that any viable $f(R)$ models must satisfy \cite{hu,sc,pogo,naniai,nunez}.

\begin{enumerate}
\item[a)] To prevent the sign change of the effective Newton's constant: $G_{eff}=\frac{G}{F(R)},~F(R)$ should be positive for all finite $R.$ Also in microscopic level, it forbids the graviton to turn to ghost models \cite{nunez}.

\item[b)] The existence of a stable high-curvature regime i.e. matter dominated universe implies $\frac{dF}{dR}>0$ for higher values of $R.$ Quantum mechanically it ensures that scalaron models are not Tachyonic.

\item[c)] $F(R)<1$ due to tight constraints from big bang nucleosynthesis and cosmic microwave background.

\item[d)] Recent galaxy formation surveys have constrained $|F(R)-1|$ to be smaller than $10^{-6}.$ As galaxy formation has not yet being studied in $F(R)$ model using $N-$body simulation so the above statement is yet to be confirmed. 
\end{enumerate} 
Thus for our solutions in equations (\ref{Fr1}), (\ref{Fr3}), (\ref{solu1b}), (\ref{solu1c}) and (\ref{F21}) the above constraints restricts the arbitrary parameters as:
\begin{center}
\begin{tabular}{|c| l |c|p{5cm} }
    \hline
    \textbf Sl.No. &{Solution for $F(R)$} &\textbf{Parameter restrictions } \\ \hline
   1.&$\frac{2R_{0}}{3R_{1}^{1/2}}I_{2}\left(2\sqrt{\frac{R_{0}}{3R_{1}^{1/2}R}}\right)$ & $\sqrt{\frac{R_{0}}{3R_{1}^{\frac{1}{2}}}}<0$ \\ \hline
   2.&$\hat{C}_{1}\left(\frac{1}{R}\right)^{\alpha_{1}}+\hat{C_{2}}\left(\frac{1}{R}\right)^{-|\alpha_{2}|}$&$0<\hat{C_{1}}<min\left\lbrace \hat{C_{2}}\frac{|\alpha_{2}|}{\alpha_{1}}R^{\alpha_{1}+|\alpha_{2}|},R^{\alpha_{1}}-\hat{C_{2}}R^{\alpha_{1}+|\alpha_{2}|}\right\rbrace $ and $\hat{C}_{1,2}>0$ \\  &$\alpha_{1}=\frac{3+2m_{1}}{4},~\alpha_{2}=\frac{3-2m_{2}}{4}$& \\ &$\hat{C}_{1,2}=C_{1,2}\left(\frac{6R_{1}+R_{0}}{R_{1}}\right)^{\alpha_{1,2}}$ &  \\ \hline
    3.&$\frac{C_{1}}{1+2n}\left(\frac{R_{2}}{R}\right)^{\frac{2n+1}{2}}+C_{2},$ & $\frac{\vert C_{1}\vert}{1+2n}\left(\frac{R_{2}}{R}\right)^{\frac{1+2n}{2}}<C_{2}<1+\frac{\vert C_{1}\vert}{1+2n}\left(\frac{R_{2}}{R}\right)^{\frac{1+2n}{2}}$ and $C_{1}<0,~(1+2n)>0$ \\ &$n\neq-\frac{1}{2}$ & \\ \hline
   4.&$C_{1}\log\left(\sqrt{\frac{6}{R}}\right)+C_{2}$ & $-\frac{C_{1}}{2}\log\left(\frac{6}{R}\right)<C_{2}<1-\frac{C_{1}}{2}\log\left(\frac{6}{R}\right)$ and $C_{1}>0$  \\ \hline
   5.&$C_{1}\left(\frac{R_{2}}{R}\right)^{\beta_{1}}+C_{2}\left(\frac{R_{2}}{R}\right)^{\beta_{2}}$& $0<C_{1}<min\left\lbrace C_{2}\frac{|\beta_{2}|}{\beta_{1}}\left(\frac{R}{R_{2}}\right)^{\beta_{1}+|\beta_{2}|},\left(\frac{R}{R_{2}}\right)^{\beta_{1}}-C_{2}\left(\frac{R}{R_{2}}\right)^{\beta_{1}+|\beta_{2}|}\right\rbrace$ and $C_{1,2}>0$\\
   &$\beta_{1,2}=\frac{(n+1)- n_{1,2}}{4}$&\\ \hline
  \end{tabular}
\end{center}

 \vspace{1em}
{\it Energy Conditions:}
\vspace{1em}

In wormhole physics, one of the basic ingredient is the violation of null energy condition. However in modified gravity theory the situation changes due to extra terms (curvature related in $f(R)$ gravity) in the field equations. In Einstein gravity, validity of null energy condition can be interpreted geometrically through Raychaudhuri equations by the geodesic focussing theorem \cite{hawking,wald} which ensures $R_{\mu\nu}\kappa^{\mu}\kappa^{\nu}\geq 0,$ where $\kappa^{\mu}$ is a null vector tangent to the null geodesics. Thus in turn from Einstein field equations we have the null energy condition. However, in modified gravity theories if the field equations are written as $G_{\mu\nu}=T_{\mu\nu}^{(eff)}$ then focusing of congruence of null geodesics will imply $T_{\mu\nu}^{(eff)}\kappa^{\mu}\kappa^{\nu}\geq 0$ \cite{albareti1, albareti2,albareti3}. For wormhole configuration in modified gravity theory with violation of energy conditions this would imply $T_{\mu\nu}^{(eff)}\kappa^{\mu}\kappa^{\nu}<0.$ Thus in principle one can impose the condition that $T_{\mu\nu}^{(m)}\kappa^{\mu}\kappa^{\nu}\geq 0$ for normal matter threading the wormhole. In $f(R)$ modified gravity theories we therefore infer that it is the higher order curvature term that is interpreted as the gravitational fluid sustaining the non-standard wormhole geometries. For the present wormhole geometry we have considered the energy conditions graphically as follows:

The matter threading the wormhole has components given by (\ref{p1}) for case 1. We check for possible weak energy conditions (WEC) violations related to both pressures. $F(R)$ solutions as obtained in equations (\ref{Fr1}), (\ref{Fr3}), (\ref{solu1b}) and (\ref{solu1c}) are used to evaluate $\rho-p_{r},~\rho-p_{t},~\rho+p_{r},~\rho+p_{t}$.

Fig 1 corresponding to (\ref{Fr1}) shows that null energy conditions (NEC) is in general satisfied for the range of parameters shown.

In Fig. 2 we have considered equation (\ref{Fr3}). Variations of NEC w.r.t. the constants $C_{1}$ and $C_{2}$ show that NEC is not violated for a large range of the considered parameter. Similar behaviour in NEC variation is observed w.r.t parameters $a_{0},~t_{0}$ and $d_{0},~r_{0}$. Also there is not much variations in the values of $\rho\pm p_{t}$ or $\rho\pm p_{r}$ with changes in the last couple of parameters.

Fig. 3 shows the variations corresponding to equation (\ref{solu1b}). First panel shows the variation for different $n$ such that $1+2n>0$. The next two panels correspond to the specific value for $n=\frac{1}{2}$. In this varying $C_{1},~C_{2}$ or $a_{0},~t_{0}$ can cause NEC violation in certain parameter ranges. Again  there is not much variations in the values of $\rho\pm p_{t}$ or $\rho\pm p_{r}$ with changes in the parameters. One must also keep in mind that $n=1/2$ is the limiting case $f(R)=R$ and actually gives wormhole solution in Einstein gravity.

Fig. 4 considers equation (\ref{solu1c}) corresponding to $n=-\frac{1}{2}.$ Here NEC is found to be violated mostly for large ranges of parameters.

\begin{figure}[htp]
\centering
\includegraphics[width=.2\textwidth]{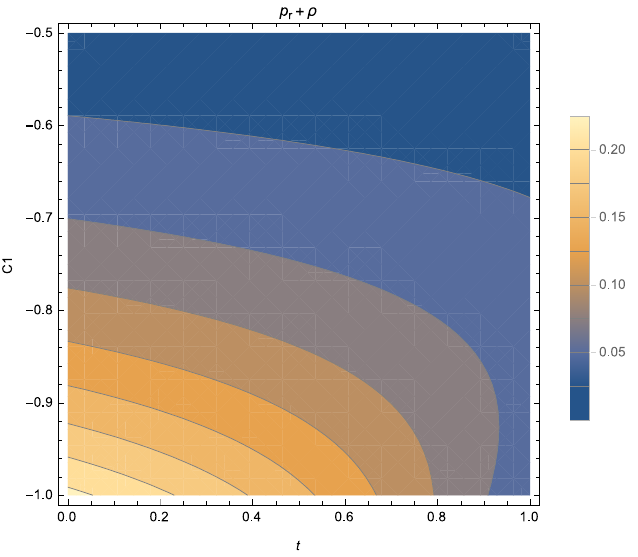}\quad
\includegraphics[width=.2\textwidth]{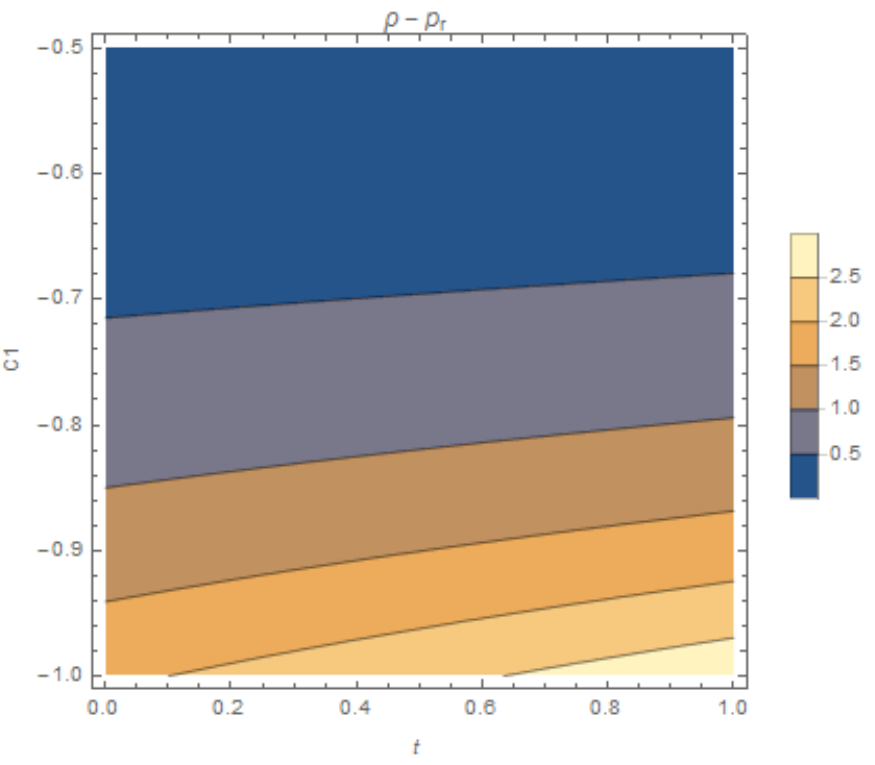}\quad
\includegraphics[width=.2\textwidth]{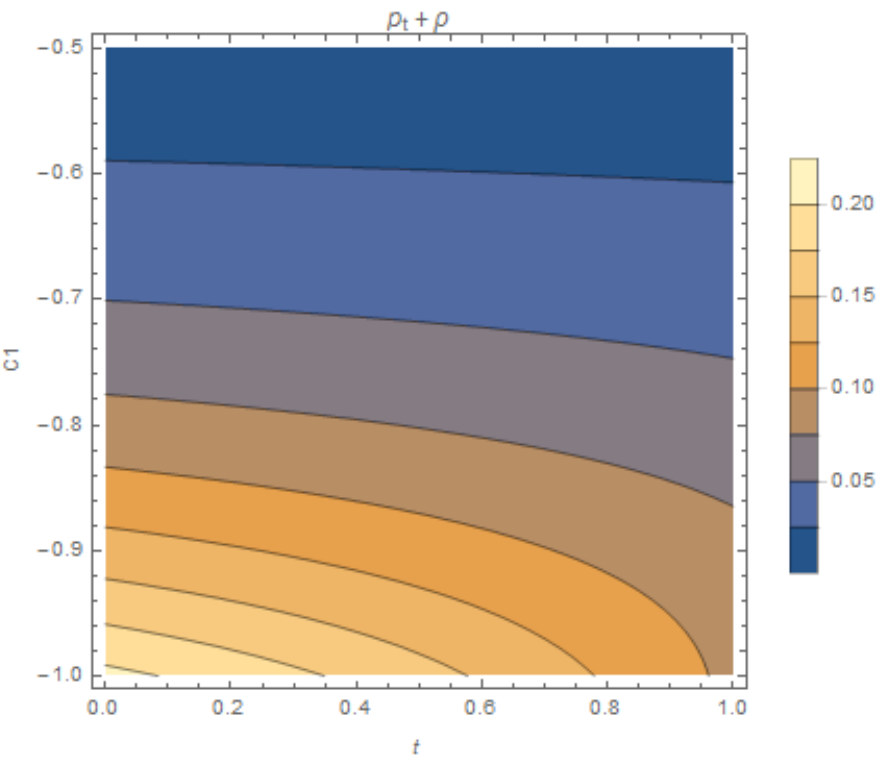}\quad
\includegraphics[width=.2\textwidth]{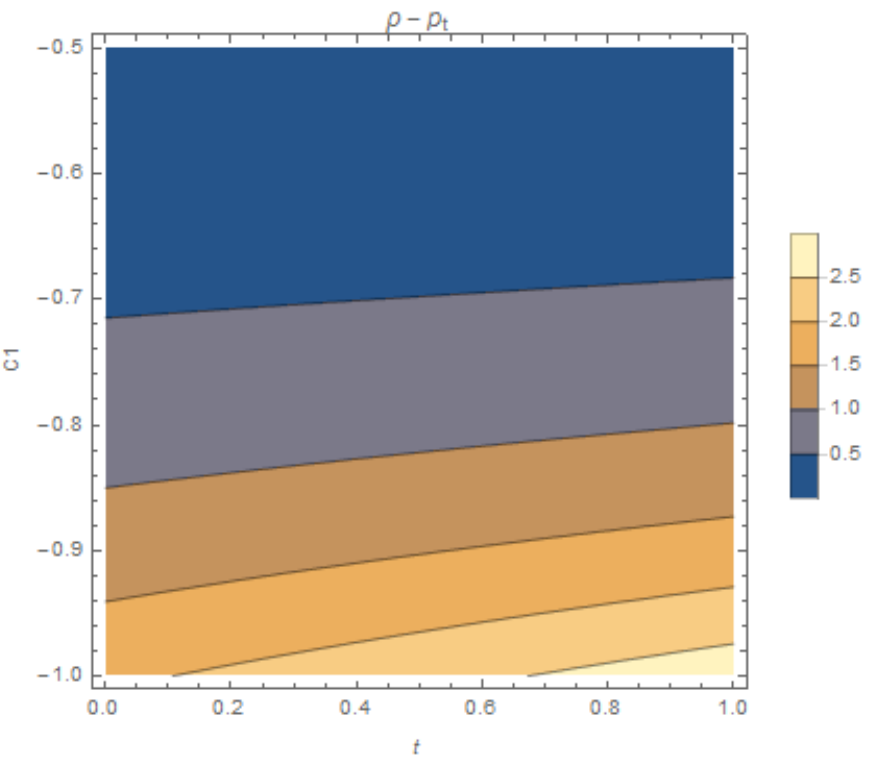}

\caption{The above panels shows the variations in $\rho\pm p_{r}$ and $\rho\pm p_{t}$ w.r.t $C_{1}$ corresponding to equation (\ref{Fr1}). Here $C_{1}=\sqrt{\frac{R_{0}}{3R^{\frac{1}{2}}_{1}}}$ is taken such that it is constrained by $f(R)$ viability conditions.}
\end{figure}

\begin{figure}[htp]
\centering
\includegraphics[width=.2\textwidth]{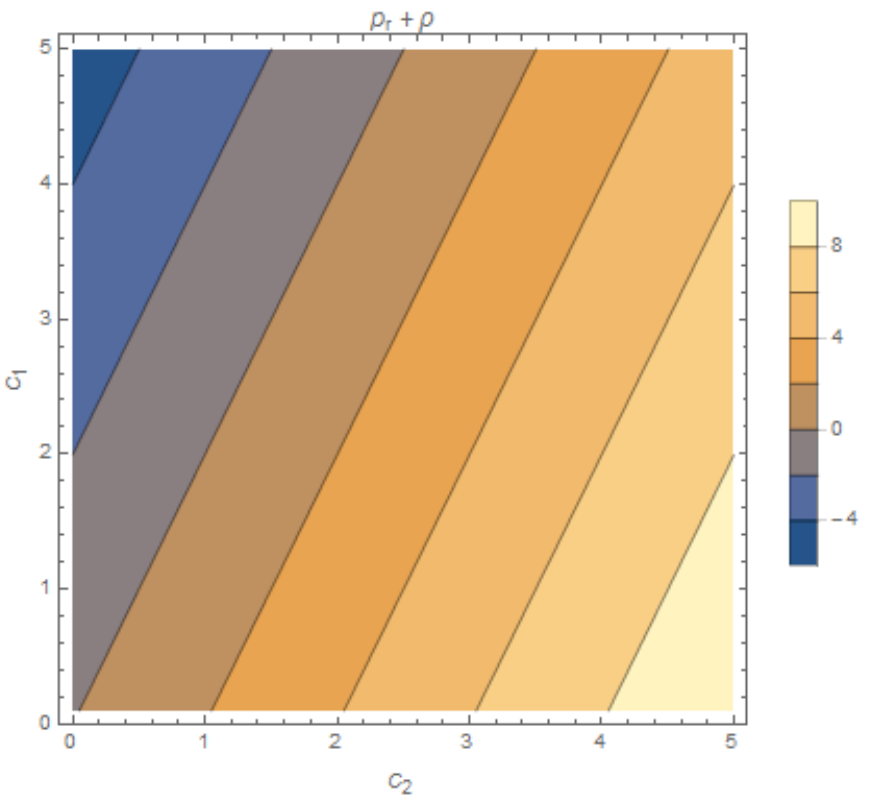}\quad
\includegraphics[width=.2\textwidth]{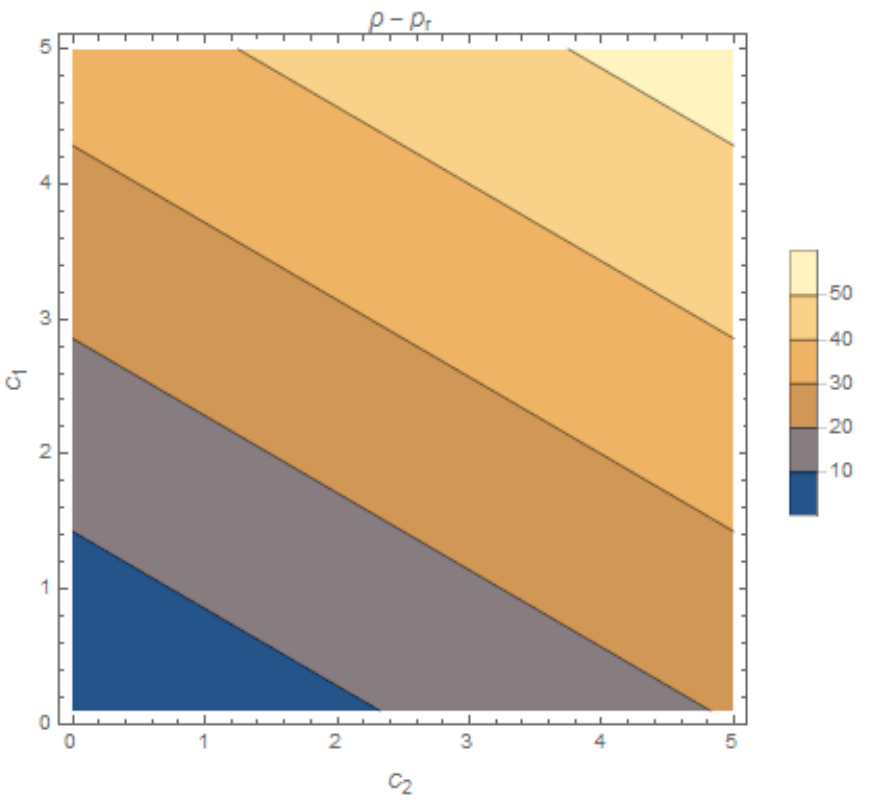}\quad
\includegraphics[width=.2\textwidth]{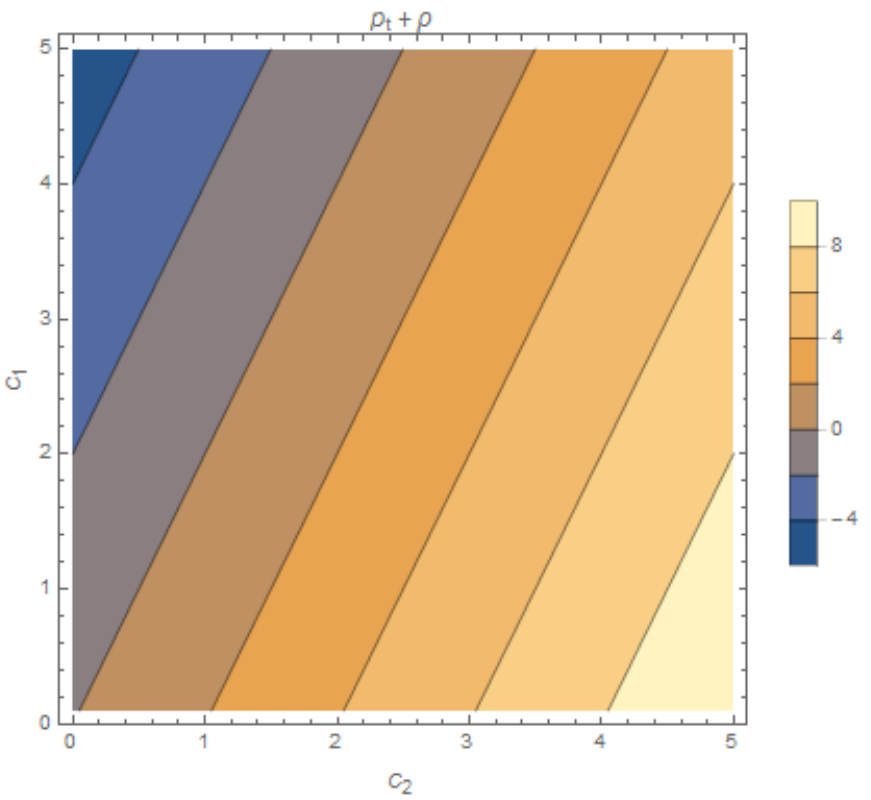}\quad
\includegraphics[width=.2\textwidth]{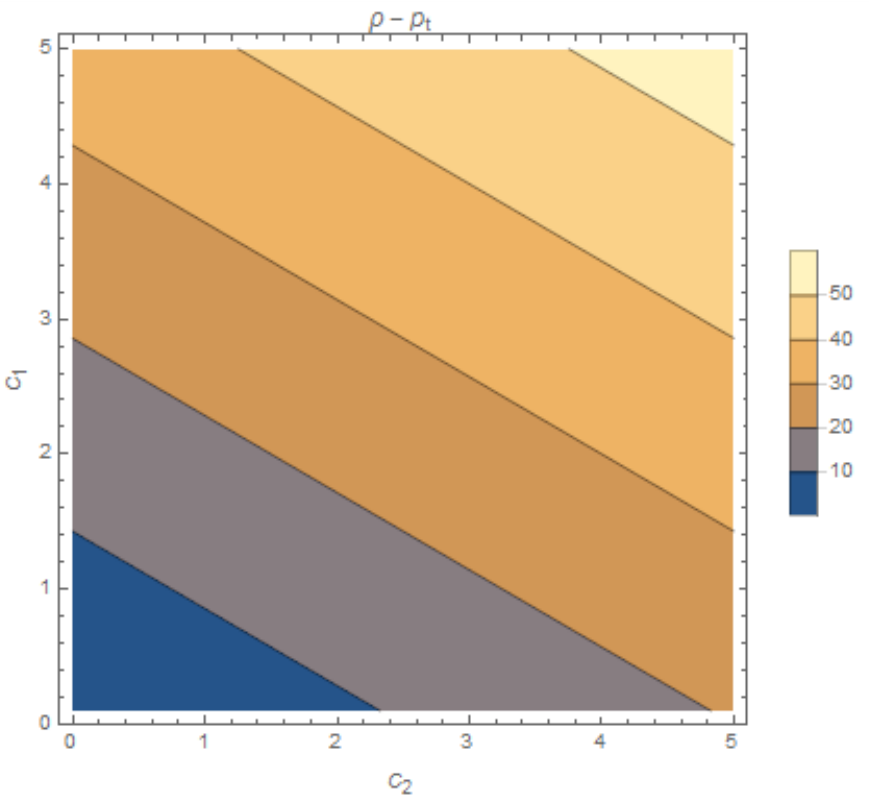}
\medskip

\includegraphics[width=.2\textwidth]{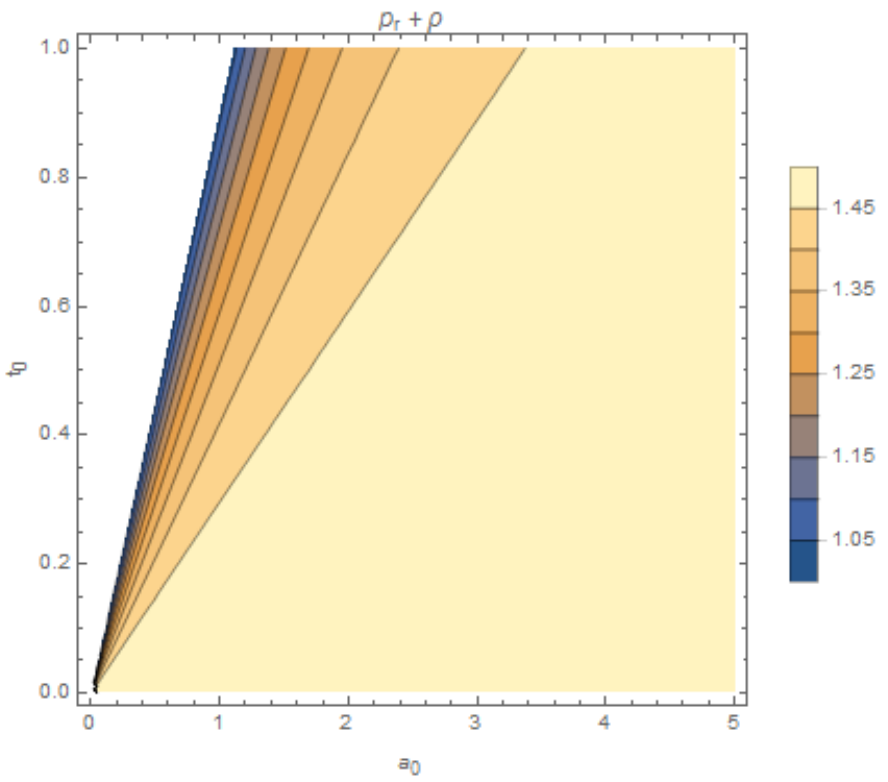}\quad
\includegraphics[width=.2\textwidth]{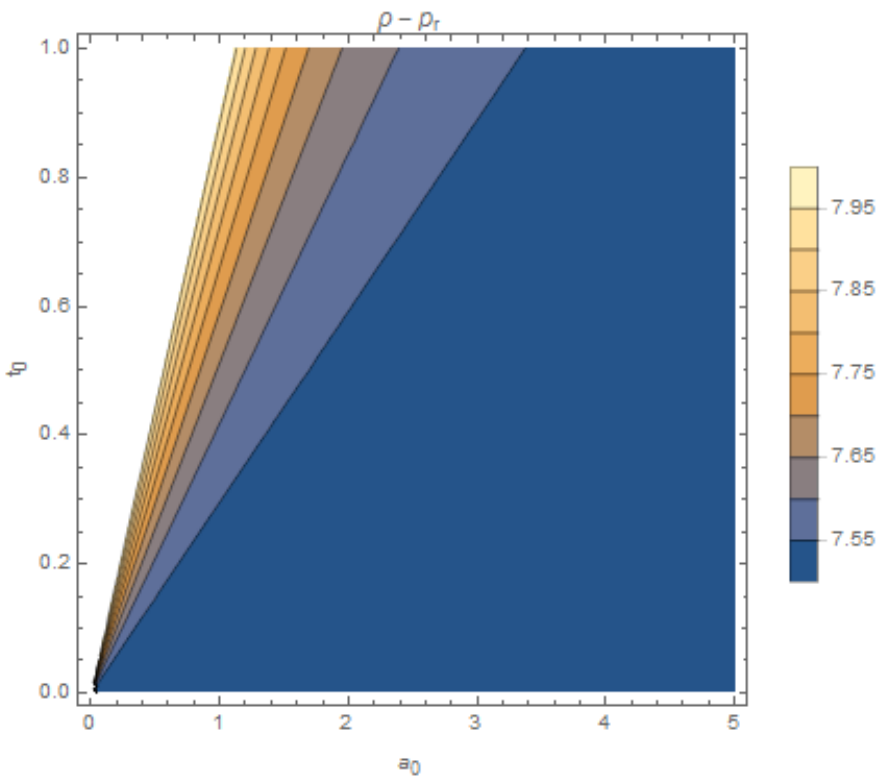}\quad
\includegraphics[width=.2\textwidth]{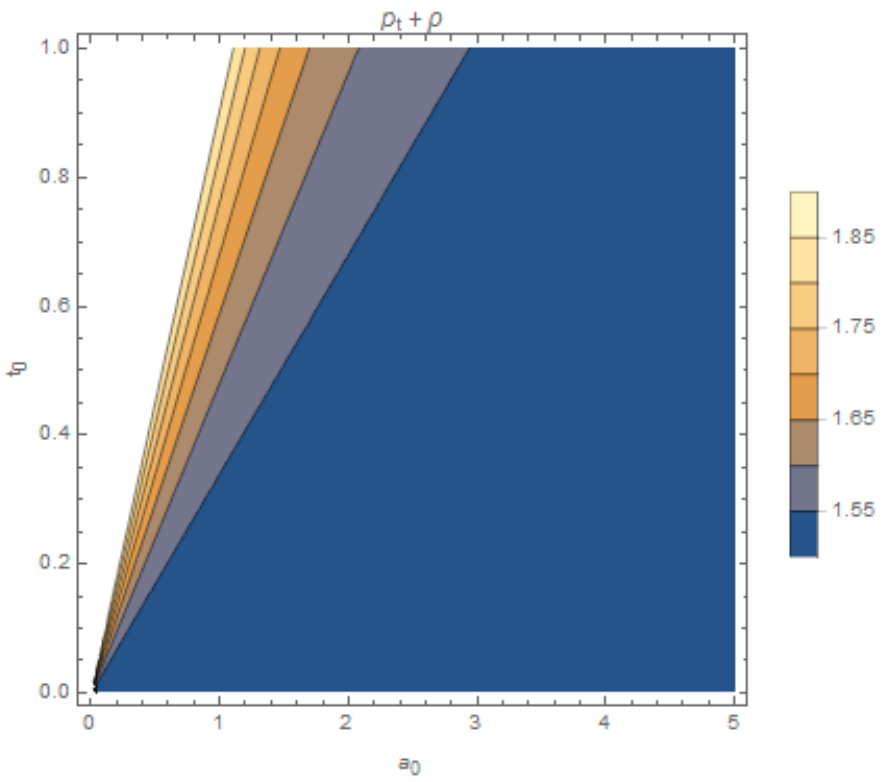}\quad
\includegraphics[width=.2\textwidth]{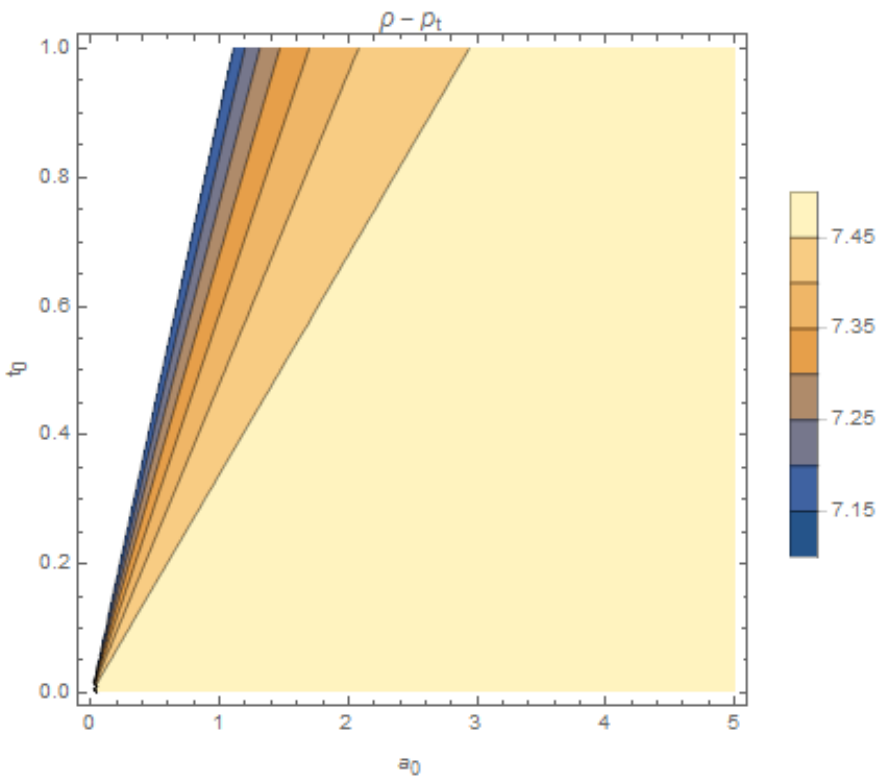}

\medskip

\includegraphics[width=.2\textwidth]{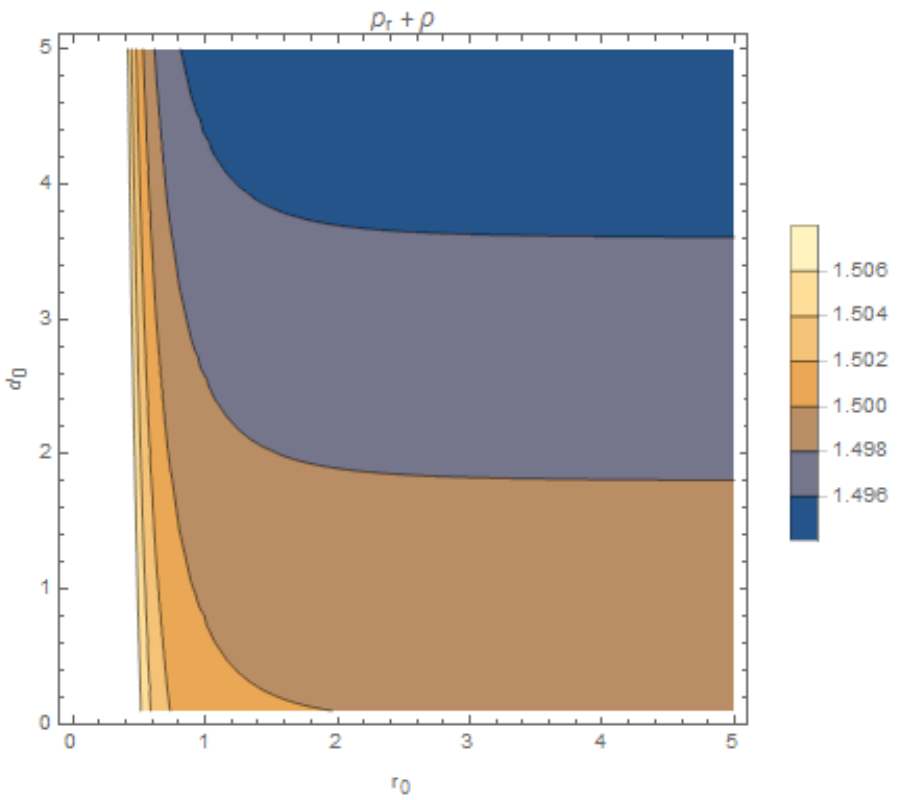}\quad
\includegraphics[width=.2\textwidth]{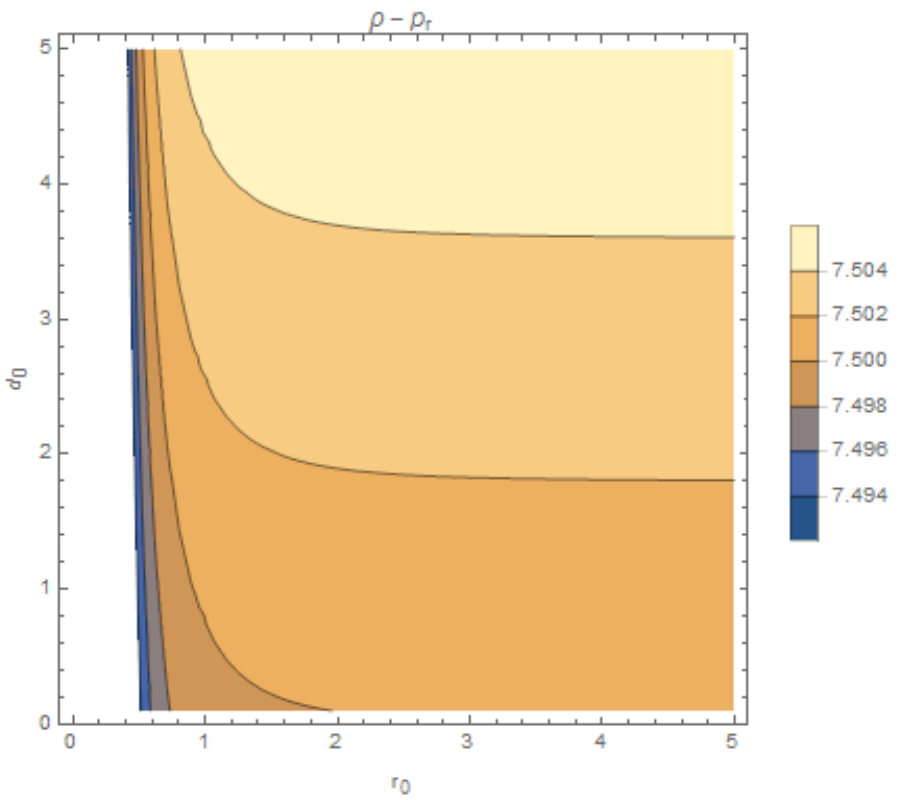}\quad
\includegraphics[width=.2\textwidth]{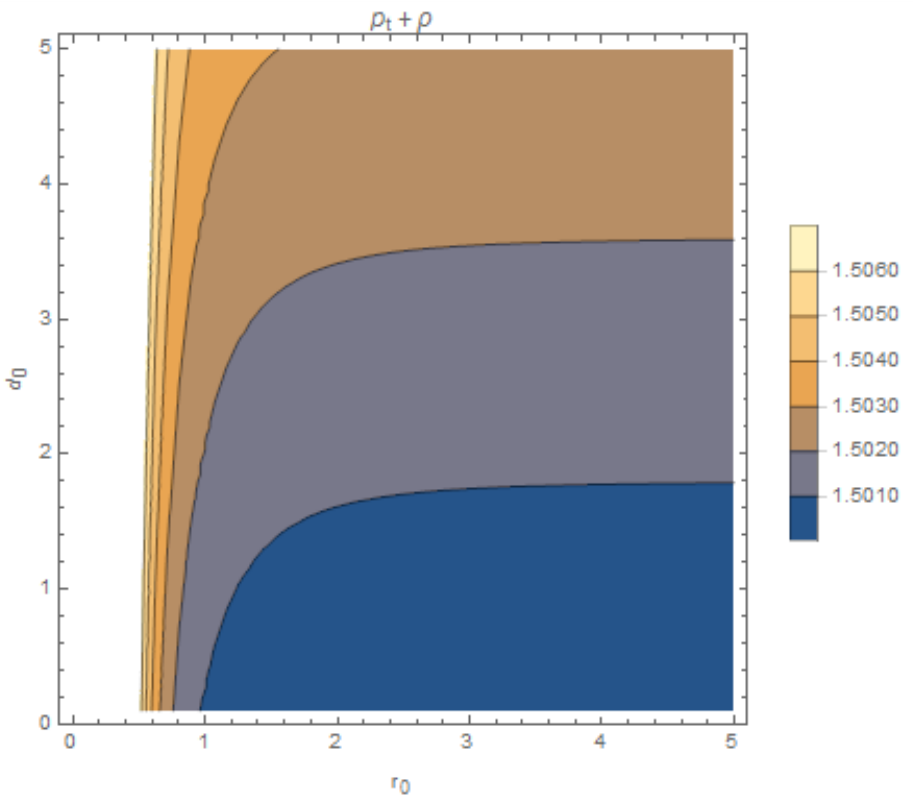}\quad
\includegraphics[width=.2\textwidth]{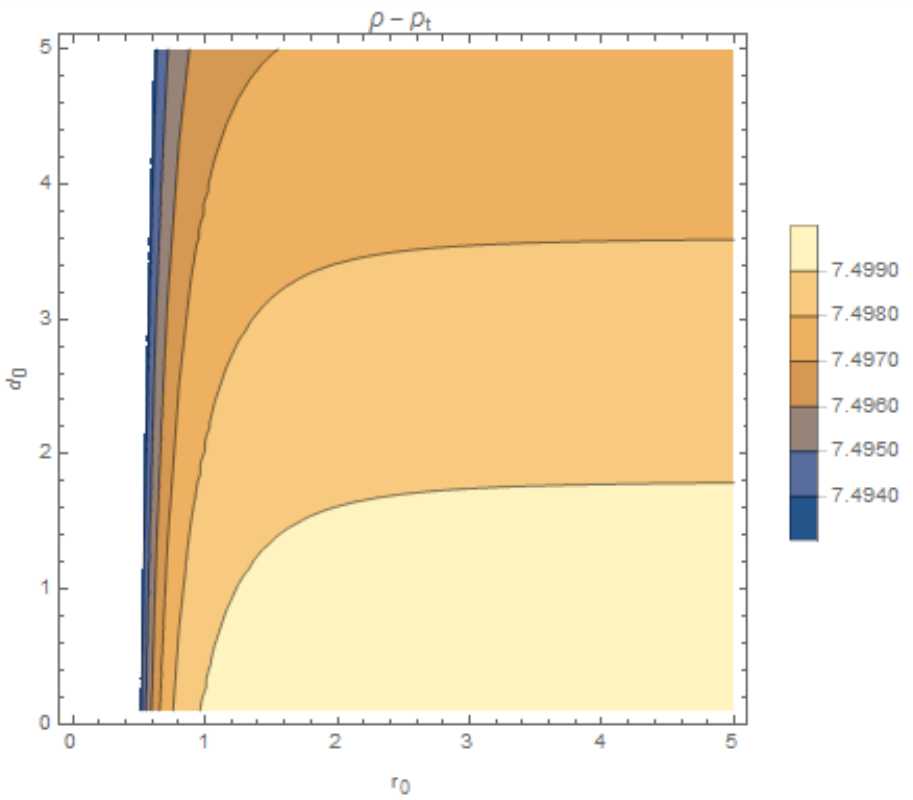}

\caption{The above panels shows the variations in $\rho\pm p_{r}$ and $\rho\pm p_{t}$ w.r.t changing parameters corresponding to equation (\ref{Fr3}). Here $C_{1},~C_{2}$ is taken with agreement to the constraints set by $f(R)$ viability conditions.}
\end{figure}

\begin{figure}[htp]
\centering
\includegraphics[width=.2\textwidth]{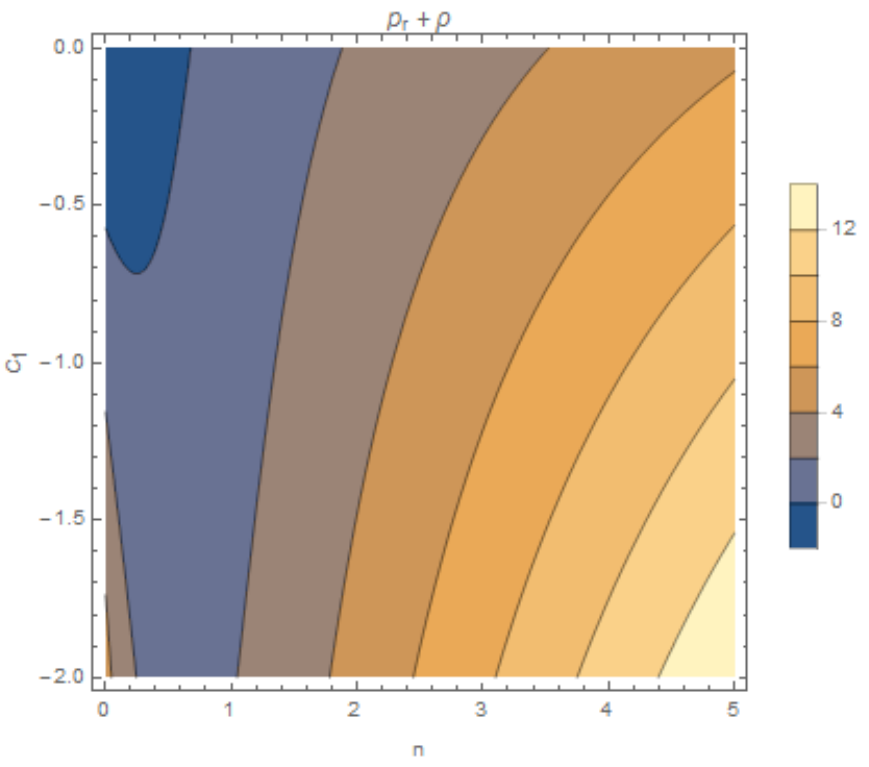}\quad
\includegraphics[width=.2\textwidth]{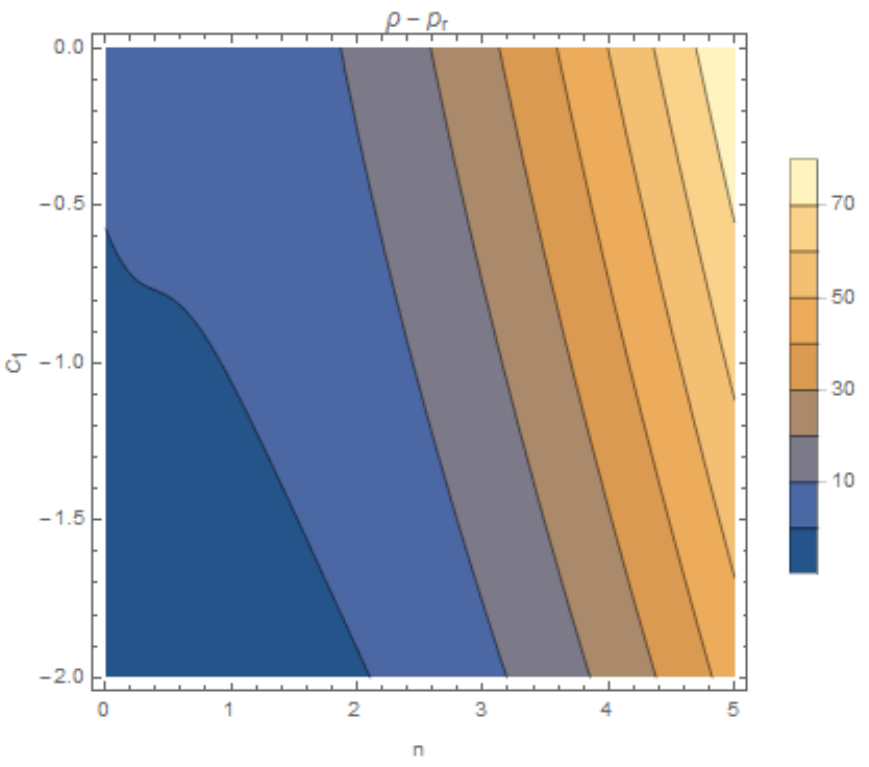}\quad
\includegraphics[width=.2\textwidth]{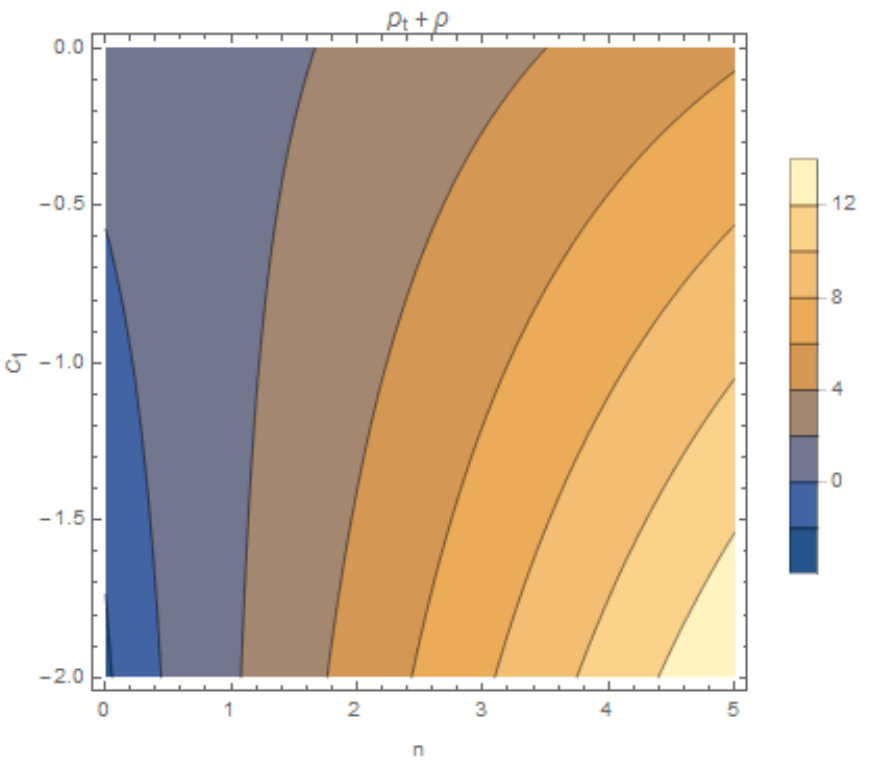}\quad
\includegraphics[width=.2\textwidth]{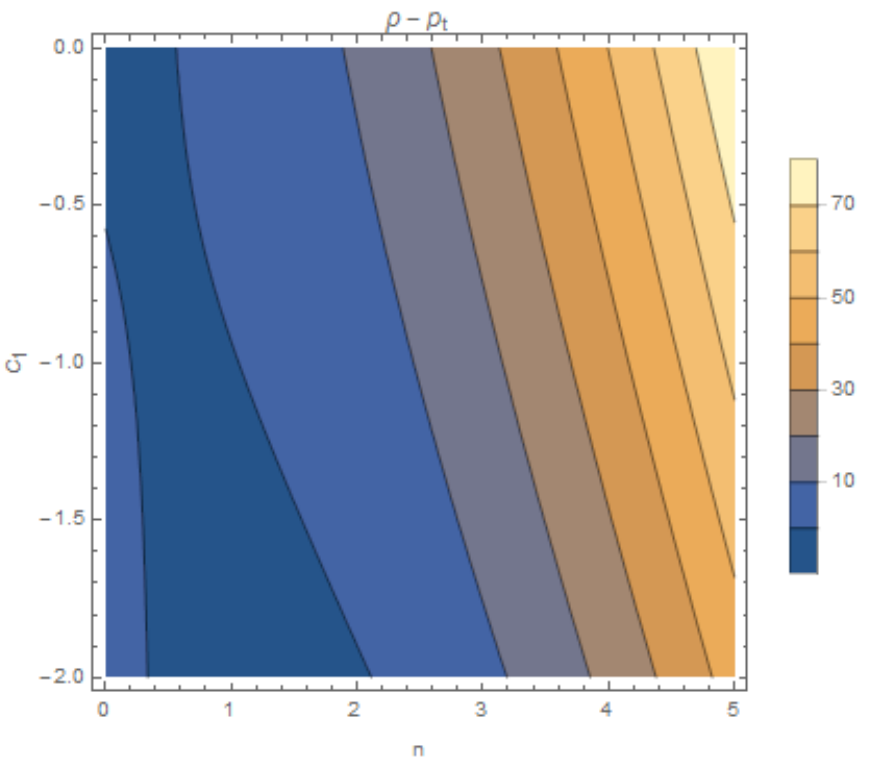}

\medskip

\includegraphics[width=.2\textwidth]{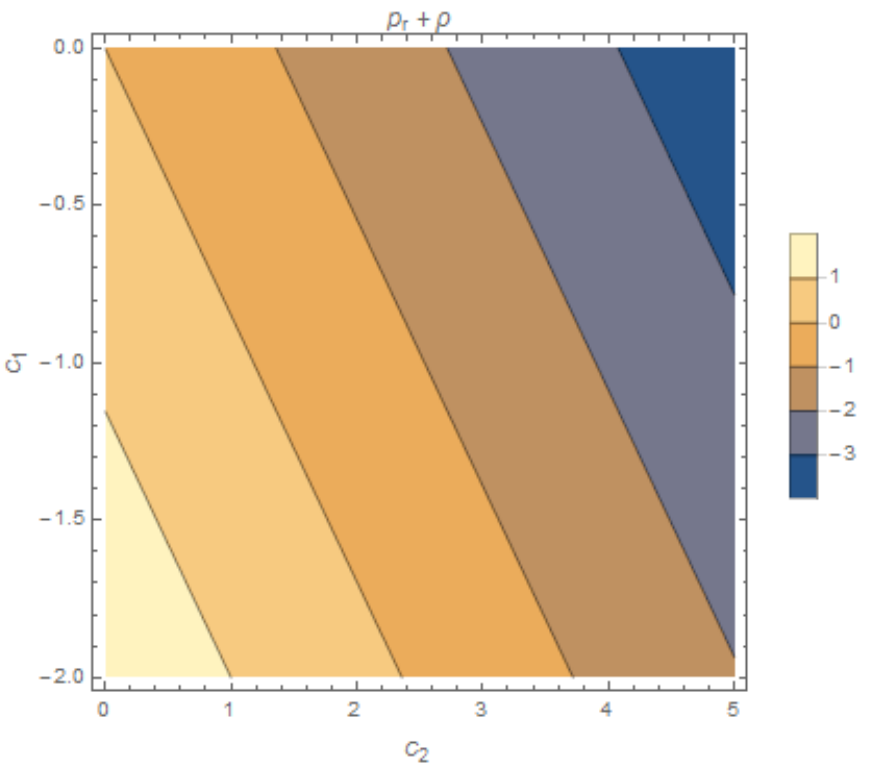}\quad
\includegraphics[width=.2\textwidth]{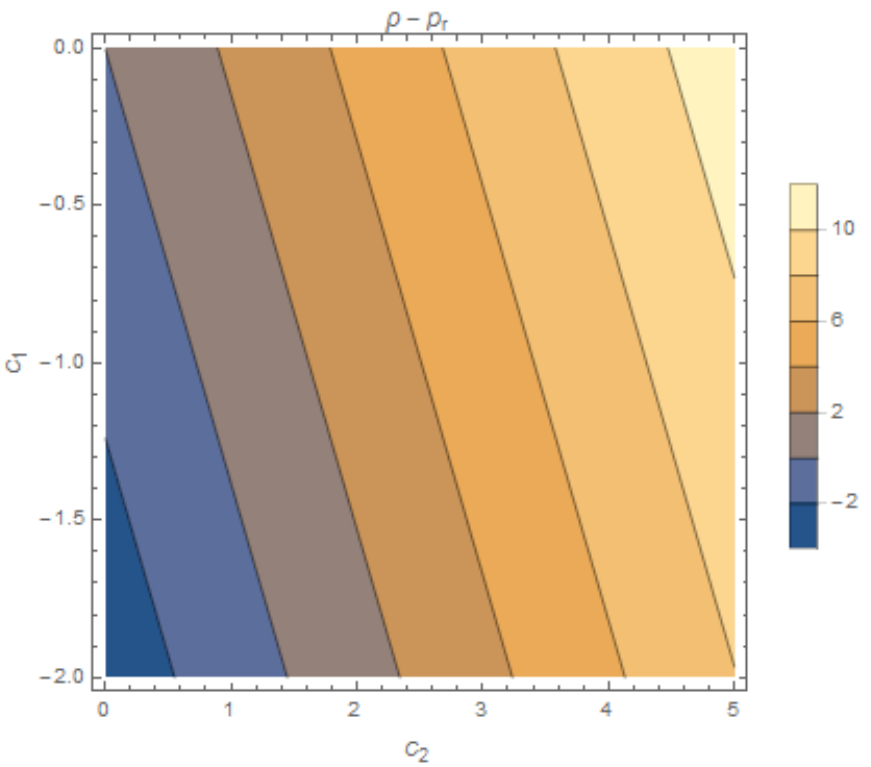}\quad
\includegraphics[width=.2\textwidth]{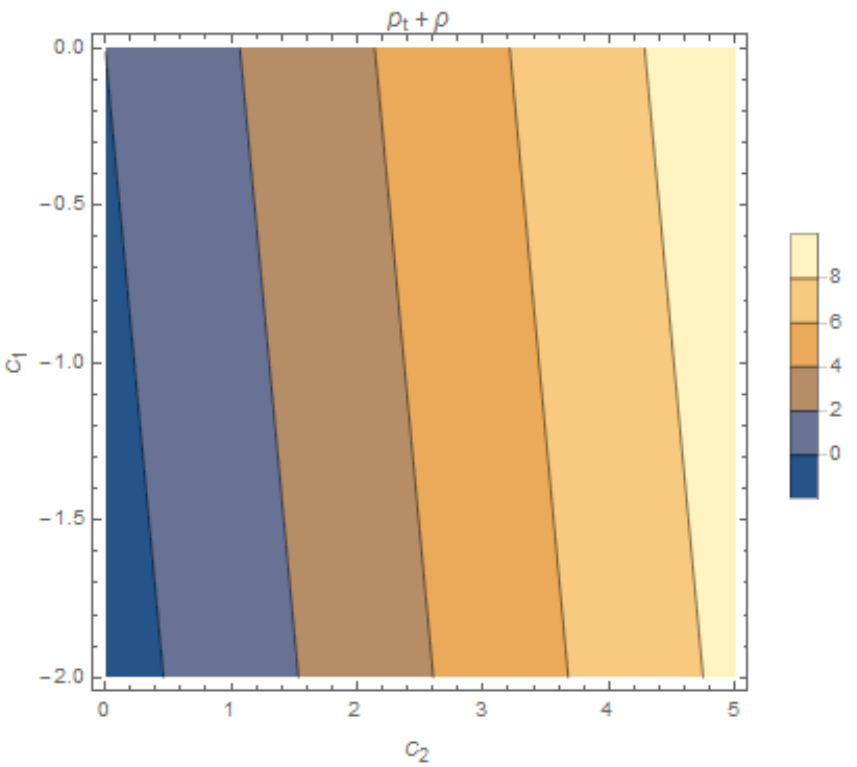}\quad
\includegraphics[width=.2\textwidth]{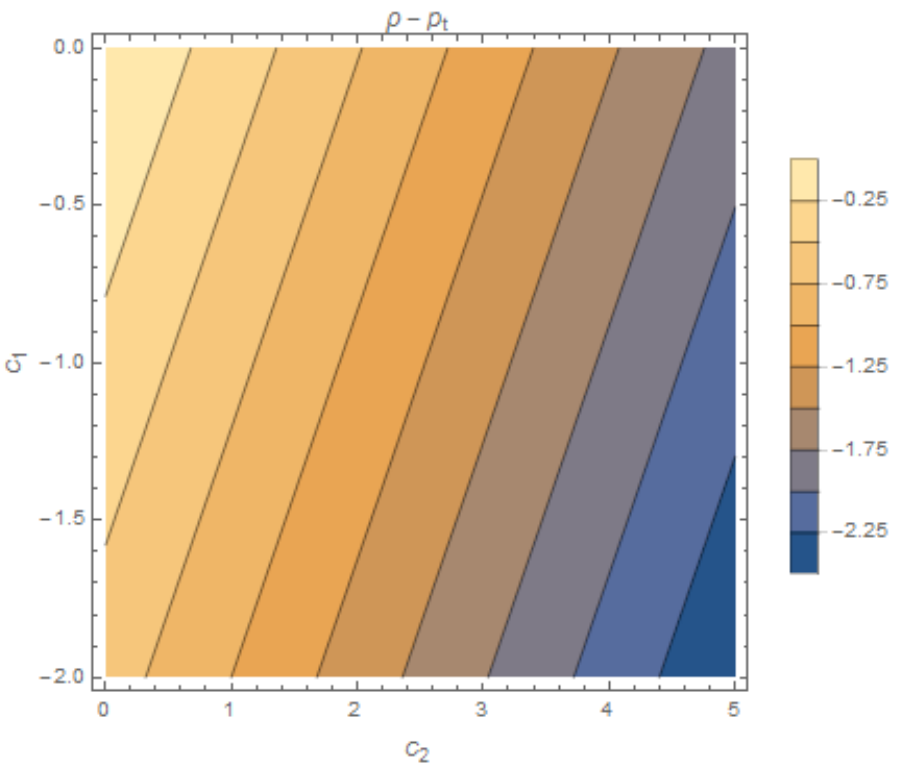}

\medskip

\includegraphics[width=.2\textwidth]{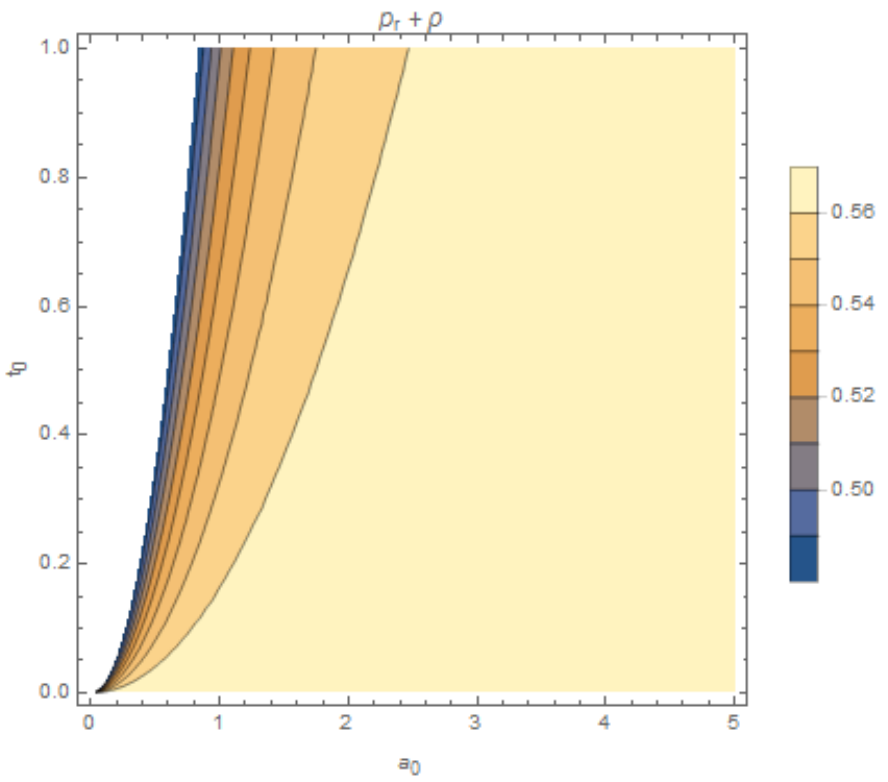}\quad
\includegraphics[width=.2\textwidth]{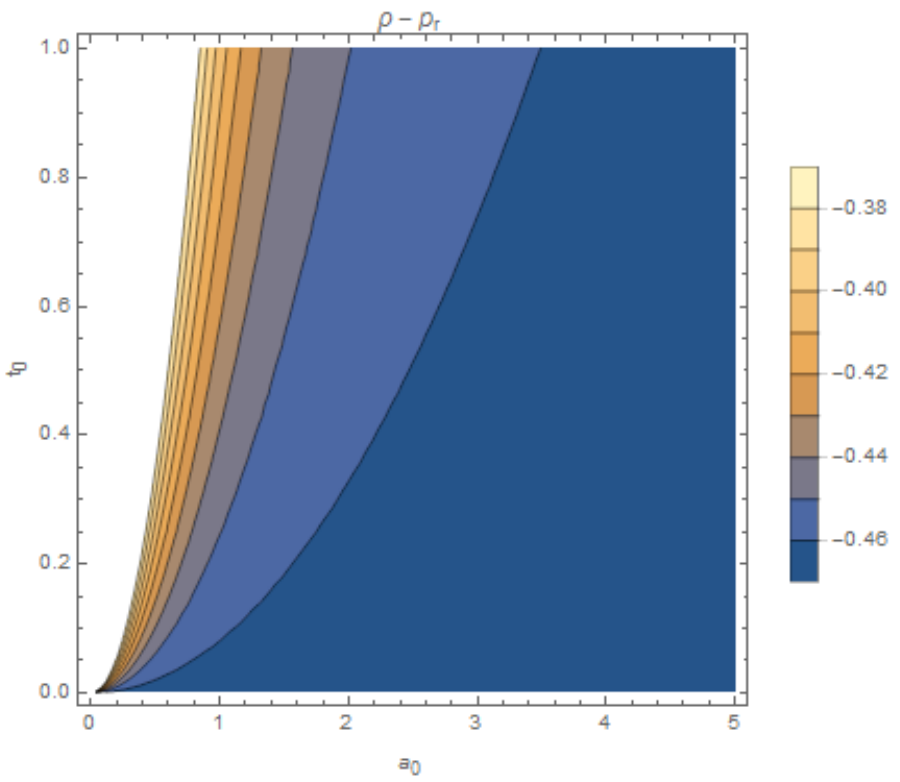}\quad
\includegraphics[width=.2\textwidth]{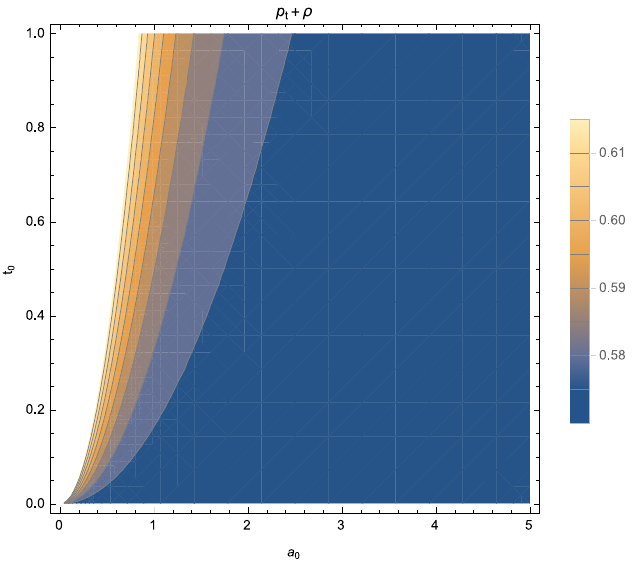}\quad
\includegraphics[width=.2\textwidth]{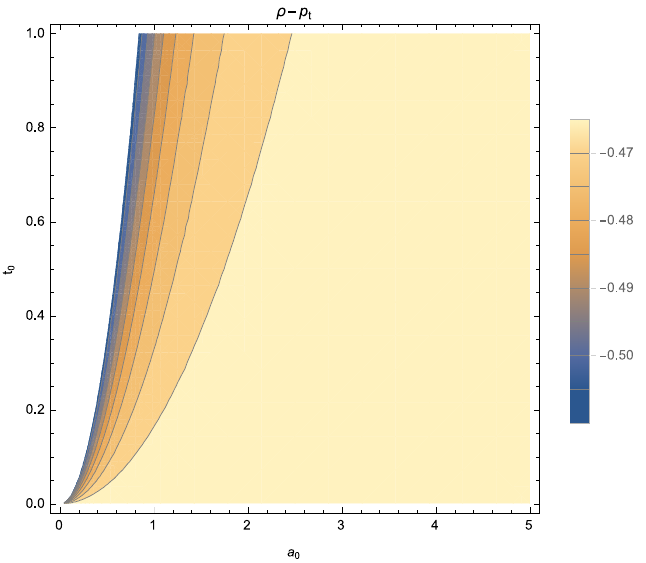}

\caption{The above panels shows the variations in $\rho\pm p_{r}$ and $\rho\pm p_{t}$ w.r.t changing parameters corresponding to equation (\ref{solu1b}). Here $C_{1},~C_{2}$ is taken with agreement to the constraints set by $f(R)$ viability conditions.}
\end{figure}

\begin{figure}[htp]
\centering
\includegraphics[width=.2\textwidth]{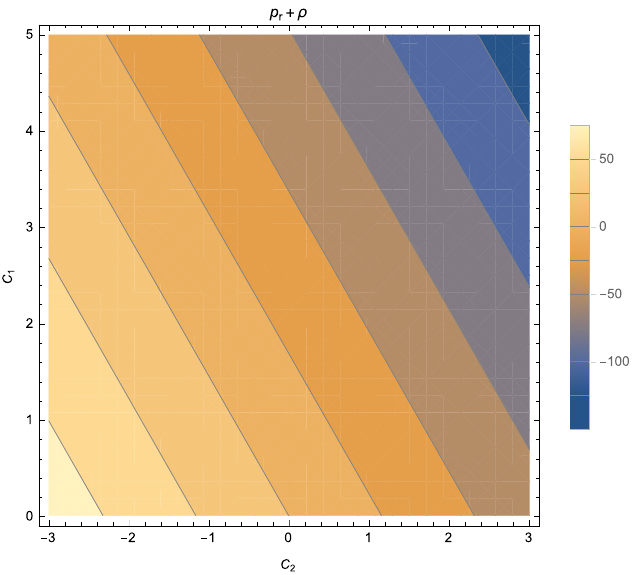}\quad
\includegraphics[width=.2\textwidth]{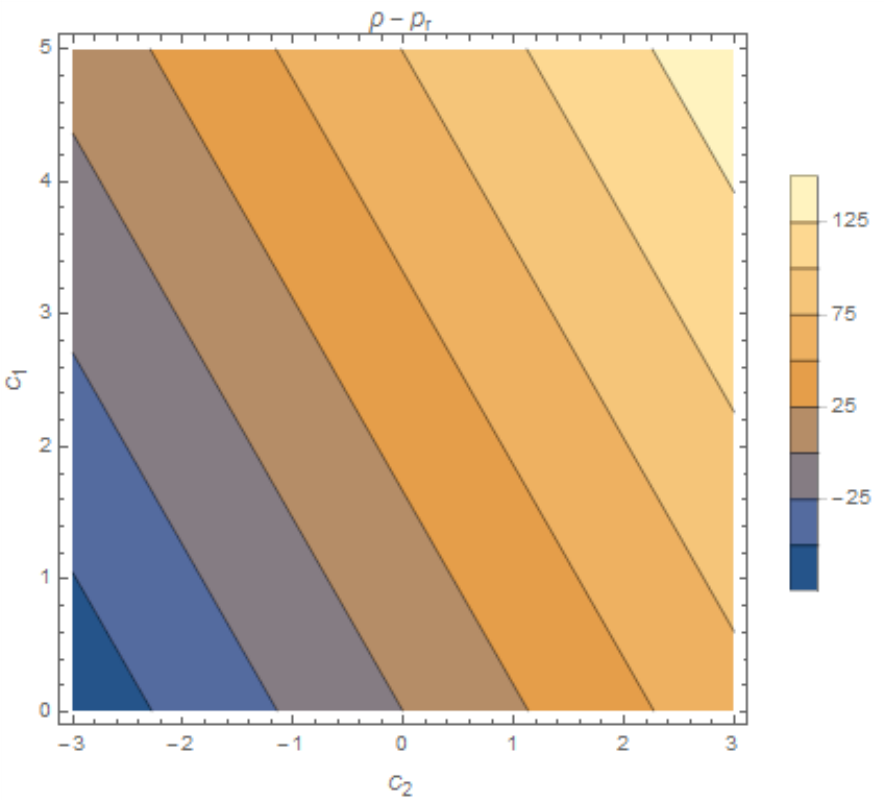}\quad
\includegraphics[width=.2\textwidth]{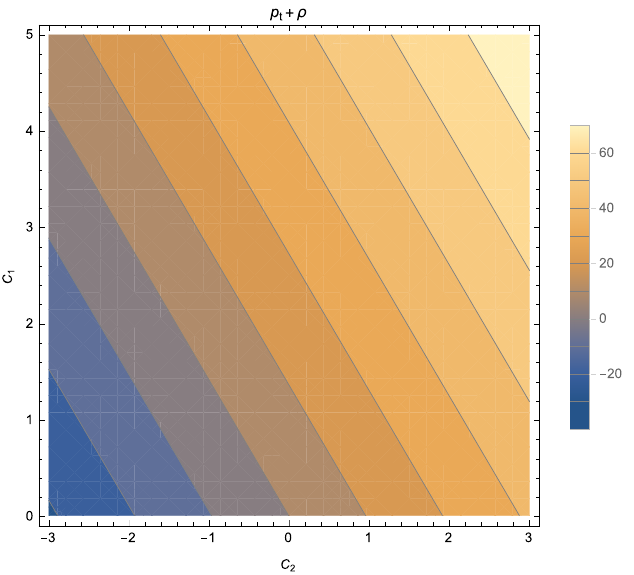}\quad
\includegraphics[width=.2\textwidth]{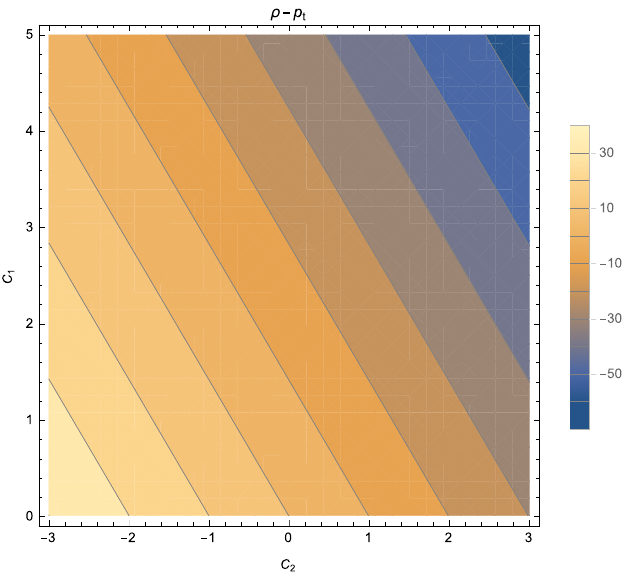}

\medskip

\includegraphics[width=.2\textwidth]{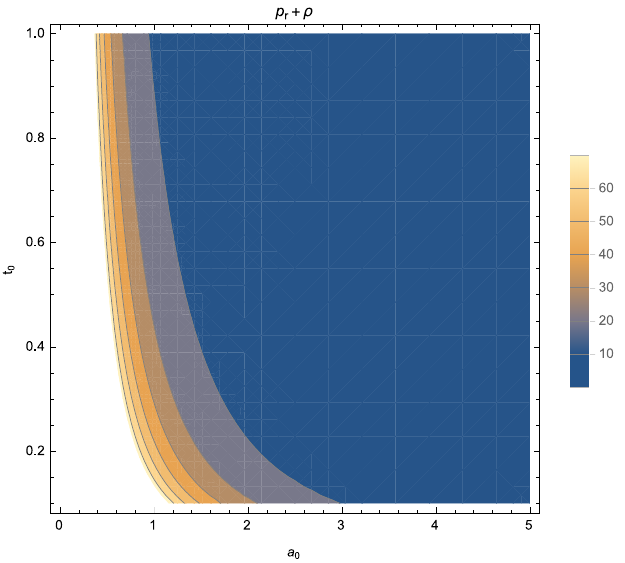}\quad
\includegraphics[width=.2\textwidth]{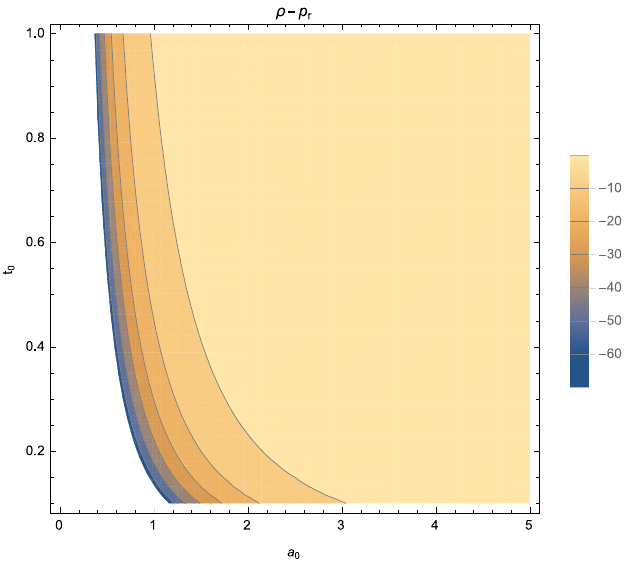}\quad
\includegraphics[width=.2\textwidth]{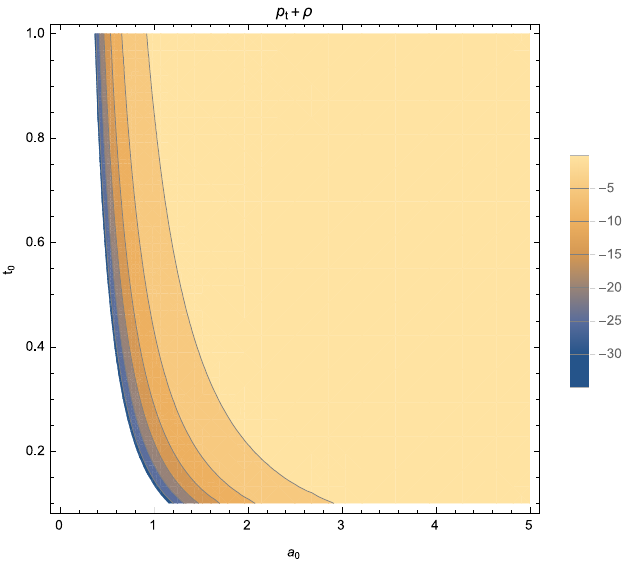}\quad
\includegraphics[width=.2\textwidth]{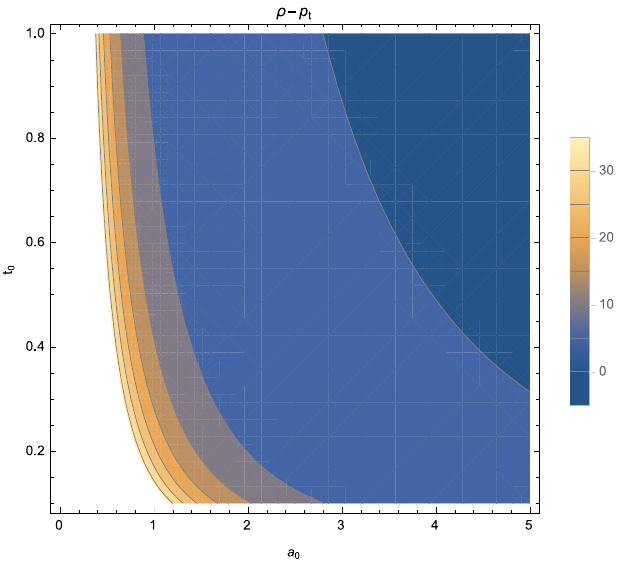}

\caption{The above panels shows the variations in $\rho\pm p_{r}$ and $\rho\pm p_{t}$ w.r.t changing parameters corresponding to equation (\ref{solu1c}). Here $C_{1},~C_{2}$ is taken with agreement to the constraints set by $f(R)$ viability conditions. }
\end{figure}

\vspace{2em}

 In the present work we have considered a simple inhomogeneous space-time metric and using the usual reconstruction technique we have found out viable $f(R)$ gravity solutions. The ensuing geometry was found to physically describe dynamical wormhole solution with variable (inhomogeneous) or constant throat radius. The resulting evolving wormhole being supported by two matter components, one being homogeneous and isotropic, the other being inhomogeneous and anisotropic. The viability of the obtained $f(R)$ solutions are examined using existing experimental and observational data. Some of the consistency conditions that are obtained for the admissibility of the $f(R)$ solutions have been presented in a tabular form above. Finally we have examined the energy conditions graphically corresponding to the various $f(R)$ solutions at the wormhole throat. From the figures one can assert that NEC in general can be obeyed by normal matter surrounding the throat in the presence of $f(R)$ gravity. Thus to conclude we state that dynamical wormholes can be made possible in $f(R)$ gravity without the violations of NEC.

\section{Acknowledgments}

SB is thankful to UGC for financial support under their Faculty Recharge Program.
SC thanks IUCAA for their research facility.


\begin{thebibliography}{}


\bibitem{ellis1} H. G. Ellis, {\it J. Math. Phys.}, {\bf 14}, 104 (1973).
\bibitem{ellis2} H. G. Ellis, {\it Gen. Relt. Grav.}, {\bf 10}, 105 (1979).
\bibitem{bronnikov} K. A. Bronnikov, {\it Acta. Phys. Poland B}, {\bf 4}, 251 (1973).
\bibitem{kodama} T. Kodama, {\it Phys. Rev. D}, {\bf 18}, 3529 (1978).
\bibitem{clement} G. Clement, {\it Gen. Relt. Grav.}, {\bf 13}, 763 (1981).
\bibitem{m+t} M. S. Morris and K. S. Thorne, {\it Am. J. Phys.}, {\bf 56}, 395 (1988).
\bibitem{visserb} M. Visser, {\it Lorentzian Wormholes: From Einstein to Hawking}, (Springer, Berlin, 1997).
\bibitem{visser1} D. Hochberg and M. Visser, {\it Phys. Rev. D}, {\bf 56}, 4745 (1997).
\bibitem{visser2} M. Visser, S. Kar and N. Dadhich, {\it Phys. Rev. Lett.}, {\bf 90}, 201102 (2003).
\bibitem{ida} D. Ida, and S. A. Hayward, {\it Phys. Lett. A}, {\bf 260}, 175 (1999).
\bibitem{fewster} C. J. Fewster and T. A. Roman, {\it Phys. Rev. D}, {\bf 72}, 044023 (2005).

\bibitem{zavalsk} O. B. Zaslavskii, {\it Phys. Rev. D}, {\bf 76}, 044017 (2007).
\bibitem{frdmn} J. L. Friedman, K. Schleich and D. M. Witt, {\it Phys. Rev. Lett.}, {\bf 71}, 1486 (1993).
\bibitem{galloway} G. J. Galloway, K. Schleich, D. M. Witt and  E. Woolgar, {\it Phys. Rev. D}, {\bf 60}, 104039 (1999).
\bibitem{visser3} M. Visser, B. Bassett and S. Liberati, {\it Nucl. Phys. Proc. Suppl.}, {\bf 88}, 267 (2000).
\bibitem{lobo1} F. S. N. Lobo, {\it Classical and Quantum Gravity Research}, (Nova Sci. Pub., p. 1-78, 2008).
\bibitem{m+t+y} M. S. Morris, K. S. Thorne and U. Yurtsever, {\it Phys. Rev. Lett.}, {\bf 61}, 1446 (1988).
\bibitem{thorne} S. W. Kim and K. S. Thorne, {\it Phys. Rev. D}, {\bf 43}, 3929 (1991).
\bibitem{visser4} M. Visser, {\it Phys. Rev. D}, {\bf 47}, 554 (1993).
\bibitem{kar1} S. Kar, {\it Phys. Rev. D}, {\bf 49}, 862 (1994).
\bibitem{kar2} S. Kar and D. Sahdev, {\it Phys. Rev. D}, {\bf 53}, 722 (1996).
\bibitem{harada1} T. Harada, H. Maeda and B. J. Carr, {\it Phys. Rev. D}, {\bf 77}, 024022, (2008).
\bibitem{harada2} H. Maeda, T. Harada and B. J. Carr, {\it Phys. Rev. D}, {\bf 77}, 024023, (2008).
\bibitem{visser6} D. Hochberg and M. Visser, {\it Phys. Rev. D}, {\bf 58}, 044021, (1998).
\bibitem{hay1} S. A. Hayward, {\it Int. J. Mod. Phys. D}, {\bf 8}, 373, (1999).

\bibitem{catl2} M. Cataldo, S. del Campo, {\it Phys. Rev. D}, {\bf 85}, 104010, (2012).
\bibitem{catl3} M. Cataldo and P. Meza, {\it Phys. Rev. D}, {\bf 87}, 064012, (2013).
\bibitem{catl4} M. Cataldo, P. Meza, P. Minning and J. Saavedra, {\it Phys. Lett. B}, {\bf 662}, 314, (2008).
\bibitem{bani} A. Banijamali and B. Fazlpour, , {\it Phys. Lett. B}, {\bf 703}, 366, (2011).
\bibitem{cai} Y-F. Cai and J. Wang, {\it Class. Quant. Grav.}, {\bf 25}, 165014, (2008).
\bibitem{pan1} S. Pan and S. Chakraborty, {\it Eur. Phys. J. C}, {\bf 73}, 2575, (2013).
\bibitem{pan2} S. Pan and S. Chakraborty, , {\it Eur. Phys. J. C}, {\bf 75}, 21, (2015).
\bibitem{visser5} N. Dadhich, S. Kar, S. Mukherjee and M. Visser, {\it Phys. Rev. D}, {\bf 65}, 064004 (2002).
\bibitem{dotti1} G. Dotti, J. Oliva and R. Troncoso, {\it Phys. Rev. D}, {\bf 75}, 024002 (2007).
\bibitem{dotti2} G. Dotti, J. Oliva and R. Troncoso, {\it Phys. Rev. D}, {\bf 76}, 064038 (2007).
\bibitem{m+n} H. Maeda and M. Nozawa, {\it Phys. Rev. D}, {\bf 68}, 024005 (2008).
\bibitem{dotti3} G. Dotti, J. Oliva and R. Troncoso, {\it Int. J. Mod. Phys. A}, {\bf 24}, 1690 (2009).
\bibitem{matulich} J. Matulich and R.Troncoso, {\it J. High Energy Phys.}, {\bf 10}, 118 (2011).
\bibitem{harko} T. Harko, F. S. N. Lobo, M. K. Mak and S. V. Sushkov, {\it Phys. Rev. D}, {\bf 87}, 067504, (2013); F. S. N. Lobo, {\it AIP Conf. Proc.}, {\bf 1458}, 447, (2011).
\bibitem{lobo} F. S. N. Lobo, M. A. Oliveira,: {\it Phys. Rev. D} {\bf 80}, 104012 (2009).
\bibitem{la} M. La Camera, {\it Phys. Lett. B}, {\bf 573}, 27, (2003).
\bibitem{furey} N. Furey and A. De Benedictis, {\it Class. Quant. Grav.}, {\bf 22}, 313, (2005); A. De Benedictis and D. Horvat, {\it Gen. Rel. Grav.}, {\bf 44}, 2711, (2012).
\bibitem{frgravit} R. Kerner, {\it Gen. Relt. Grav.}, {\bf 14}, 453, (1982); J. P. Druisseau, R. Kerner and P. Eysseric {\it Gen. Relt. Grav.}, {\bf 15}, 797, (1983); J. D. Barrow and A. C. Ottewill, {\it J. Phys. A: Math. Gen.}, {\bf 16}, 2757, (1983); T. P. Sotiriou and V. Faraoni, {\it Rev. Mod. Phys.}, {\bf 82}, 451, (2010).
\bibitem{weil} H. Weil, {\it Sitzungsberichte der Koniglich Preussischen Akademie der Wissenchaften zu Berlin, GA II}, {\bf 31}, p. 29-42, (1918), L. O'Raifeartaigh, {\it The Dawning of Gauge Theory}, (Princeton University Press, p. 24-37, 1997).
\bibitem{dono}J. F. Donoghue, {\it Gravity as an Effective Field Theory. Dispersive Techinques in Effective Field Theories}, ( Advanced School
on Effective Theories:, Almu\~{n}ecar, Granada (Spain), 25 June - 1 July 1995),  arXiv:gr-qc/9512024.  
\bibitem{starobin} A. A. Starobinsky, {\it Phys. Lett. B}, {\bf 91}, 99, (1980).
\bibitem{drkeng} S. Capozziello and M. Francaviglia, {\it Gen. Relt. Grav.}, {\bf 40}, 357, (2008); S. Capozziello, S. Carloni and A. Troisi, {\it Recent Res. Dev. Astron. Astrophys.}, {\bf 1}, 625, (2003); S. M. Carroll, V. Duvvuri, M. Trodden and M. S. Turner, {\it Phys. Rev. D}, {\bf 70}, 043528, (2004); S. Nojiri and S. D. Odinstov, {\it Int. J. Geom. Meth. Mod. Phys.}, {\bf 4}, 115, (2007). 
\bibitem{capo}S. Capozziello and M. DeLaurentis, {\it Annelen Phys.}, {\bf 524}, 545, (2012).
\bibitem{cemb}J. A. R.Cembranos, {\it Phys. Rev. Lett.}, {\bf 102}, 141301, (2009).
\bibitem{capo1}S. Capozziello, M. De Laurentis, and O. Luongo, {\it Int. J. Mod. Phys. D}, {\bf 24}, 1541002, (2015). 
\bibitem{dom} A. de la Cruz-Dombriz, A. Dobado {\it Phys. Rev. D}, {\bf 74}, 087501, (2006).
\bibitem{noj} S. Nojiri, S. D. Odintsov, A. Toporensky, P. Tretyakov, {\it Gen.Rel.Grav.}, {\bf 42}, 1997, (2010). 
\bibitem{cog1} G. Cognola, E. Elizalde, S. Nojiri, S. D. Odintsov, {\it Open Astron.J.}, {\bf 3}, 20, (2010).
\bibitem{dunb}P. K. S. Dunsby, E. Elizalde, R. Goswami, S. Odintsov, D. Saez-Gomez, {\it Phys. Rev. D}, {\bf 82}, 023519, (2010).
\bibitem{carl}S. Carloni, R. Goswami, Peter K.S. Dunsby, {\it Class. Quant. Grav.}, {\bf 29}, 135012, (2012)

\bibitem{catl}Mauricio Cataldo, Luis Liempi, Pablo Rodriguez, {\it Phys. Lett. B}, {\bf 757}, 130, (2016).
\bibitem{cog} G. Cognola, E. Elizalde, S. Nojiri, S. D. Odintsov, L. Sebastiani, S. Zerbini, {\it Phys. Rev. D}, {\bf 77}, 046009, (2008).
\bibitem{lind} E. V. Linder, {\it Phys. Rev. D}, {\bf 80}, 123528, (2009).
\bibitem{nojiri} S. Nojiri and S. D. Odintsov, {\it Phys. Lett. B}, {\bf 576}, 5, (2003).
\bibitem{vassi} D. V. Vassilevich, {\it Phys. Rept.}, {\bf 388}, 279, (2003).
\bibitem{nojiri1} S. Nojiri and S. D. Odintsov, {\it Phys. Rev. D}, {\bf 68}, 123512, (2003).
\bibitem{sotirio} T. P. Sotiriou, {\it Class. Quant. Gravit.}, {\bf 23}, 1253, (2006).
\bibitem{sotirio1} T. P. Sotiriou, {\it Phys. Rev. D}, {\bf 73}, 063515, (2006)
\bibitem{whitt} B. Whitt, {\it Phys. Lett. B}, {\bf 145}, 176, (1984).
\bibitem{maeda} K. Maeda, {\it Phys. Rev. D}, {\bf 39}, 3159, (1989).
\bibitem{chiba} T. Chiba, {\it Phys. Lett. B}, {\bf 575}, 1, (2003).
\bibitem{magnano} G Magnano and L. M. Sokolowski, {\it Phys. Rev. D}, {\bf 50}, 5039, (1994).
\bibitem{khoury} J. Khoury and A. Weltman, {\it Phys. Rev. Lett.}, {\bf 93}, 171104, (2004); J. Khoury and A. Weltman, {\it Phys. Rev. D}, {\bf 69}, 044026, (2004).
\bibitem{hu} W. Hu and I. Sawicki, {\it Phys. Rev. D}, {\bf 76}, 064004, (2007).
\bibitem{sc} S. Chakraborty and S. SenGupta, {\it Phys. Rev. D}, {\bf 90}, 047901, (2014).
\bibitem{pogo} L. Pogosian and A. Silvestri, {\it Phys. Rev. D}, {\bf 77}, 023503, (2008).
\bibitem{naniai} H. Naniai, {\it Prog. Theo. Phys.}, {\bf 49}, 165, (1973). 
\bibitem{nunez} A. Nunez and S. Solganik, arXiv:hep-th/0403159, (2004).  
\bibitem{hawking} S. W. Hawking and G. F. R. Ellis, {\it The large scale structure of space-time}, (Cambridge University Press, Cambridge, U.K., 1973).
\bibitem{wald} R. M. Wald, {\it General Relativity}, (University of Chicago
Press, Chicago, U.S.A., 1984).
\bibitem{albareti1} F.D. Albareti, J.A.R. Cembranos and A. de la Cruz-Dombriz, {\it J. Cosmol. Astropart. Phys.}, {\bf 1212}, 020, (2012).
\bibitem{albareti2} F.D. Albareti, J.A.R. Cembranos, A. de la Cruz-Dombriz, and A. Dobado, {\it J. Cosmol. Astropart. Phys.}, {\bf 1307}, 009, (2013).
\bibitem{albareti3} F.D. Albareti, J.A.R. Cembranos, A. de la Cruz-Dombriz, and A. Dobado, {\it J. Cosmol. Astropart. Phys.}, {\bf 1403}, 012, (2014).
\end{thebibliography}
\end{document}